\documentclass{article}

\usepackage[a4paper, total={6in, 8in}]{geometry}
\usepackage{enumerate}
\usepackage{mathtools}
\usepackage{amsfonts}
\usepackage{amsbsy}
\usepackage{amssymb}
\usepackage{amsthm}
\usepackage{booktabs}
\usepackage[table]{xcolor}
\usepackage{graphicx}
\usepackage[switch]{lineno}
\usepackage{mathtools}
\usepackage{algorithm}
\usepackage[noend]{algpseudocode}
\usepackage{amsmath}
\usepackage{calrsfs}
\DeclareMathAlphabet{\pazocal}{OMS}{zplm}{m}{n}
\usepackage{amsfonts}
\usepackage{stackengine}
\stackMath
\usepackage{siunitx}
\usepackage{diagbox}
\usepackage{color}
\usepackage{subcaption}
\usepackage{hyperref}
\usepackage{soul}
\usepackage{comment}
\usepackage{multirow}

\def\bn{\mathbf{n}}
\def\bv{\mathbf{v}}

\def\bt{\mathbf{t}}

\usepackage{cleveref}
\Crefname{figure}{Fig.}{Figs.}
\Crefname{equation}{Eq.}{Eqs.}
\Crefname{table}{Tab.}{Tabs.}
\Crefname{appendix}{App.}{Apps.}
\Crefname{theorem}{Prop.}{Props.}
\Crefname{section}{Sec.}{Secs.}

\begin{document}
\title{The de Rham-Hodge analysis and modeling of biomolecules }
\author{Rundong Zhao$^1$, Menglun Wang$^2$, Yiying Tong$^1$\footnote{
		Corresponding author:		Email: ytong@msu.edu}, and Guo-Wei Wei$^{2,3,4}$\footnote{
		Corresponding author.		Email: wei@math.msu.edu} \\
$^1$ Department of Computer Science and Engineering,\\
Michigan State University, MI 48824, USA.\\
$^2$ Department of Mathematics, \\
Michigan State University, MI 48824, USA.\\
$^3$ Department of Electrical and Computer Engineering,\\
Michigan State University, MI 48824, USA. \\
$^4$ Department of Biochemistry and Molecular Biology,\\
Michigan State University, MI 48824, USA. \\
}
\date{\today} 

\maketitle

\begin{abstract}
Biological macromolecules have intricate structures that underpin their biological functions. Understanding their structure-function relationships remains a challenge due to their structural complexity and functional variability.  Although de Rham-Hodge theory, a landmark of 20th Century's mathematics, has had a tremendous impact on mathematics and physics, it has not been devised for macromolecular modeling and analysis. In this work, we introduce de Rham-Hodge theory as a unified paradigm for analyzing the geometry, topology, flexibility, and natural modes of biological macromolecules. Geometric characteristics and topological invariants are obtained either from the Helmholtz-Hodge decomposition of the scalar, vector and/or tensor fields of a macromolecule or from the spectral analysis of various Laplace-de Rham operators defined on the molecular manifolds. We propose   Laplace-de Rham spectral based models for predicting macromolecular flexibility.  We further construct a Laplace-de Rham-Helfrich operator for revealing cryo-EM natural modes. Extensive experiments are carried out to demonstrate that the proposed de Rham-Hodge paradigm is one of the most versatile tools for the multiscale modeling and analysis of biological macromolecules and subcellular organelles.  Accurate, reliable and topological structure-preserving algorithms for implementing discrete exterior calculus (DEC) have been developed to facilitate the aforementioned modeling and analysis of biological macromolecules. The proposed de Rham-Hodge paradigm has potential applications to subcellular organelles and the structure construction from medium or low-resolution cryo-EM maps, and functional predictions from massive biomolecular datasets.

\end{abstract}
Key words: Algebraic topology; Differential geometry, de Rham-Hodge theory, macromolecular flexibility, Macromolecular natural mode, Cryo-EM analysis



\section{Introduction}

One of the most amazing aspects of biological science is the intrinsic structural complexity of biological macromolecules and its associated function. The understanding of how changes in macromolecular structural complexity alter their function remains one of the most challenging issues in biophysics, biochemistry, structural biology, and molecular biology.  This understanding depends crucially on our ability to model three-dimensional (3D) macromolecular shapes from original experimental data and to extract geometric and topological information from the architecture of molecular structures. Very often, macromolecular functions depend not only on native structures but also on nascent, denatured or unfolded states. As a result, understanding the structural instability, flexibility, and collective motion of macromolecules are of vital importance.  Structural bioinformatics searches patterns among diverse geometric, topological,  instability and dynamic features to deduce macromolecular function. Therefore, the development of efficient and versatile computational tools for extracting macromolecular geometric characteristics, topological invariants, instability spots, flexibility traits, and natural modes is a key to infer their functions, such as binding affinity, folding, folding stability change upon mutation, reactivity, catalyst efficiency,  allosteric effects, etc.

Geometric modeling and characterization of macromolecular 3D shapes have been an active research topic for many decades. Surface models not only provide a visual basis for understanding macromolecular 3D shapes and but also bridge the gap between experimental data and theoretical modeling, such as generalized Born and Poisson-Boltzmann models for biomolecular electrostatics \cite{NKWH07,ZYu:2008}. A space-filling model with van der Waals spheres was introduced by  Corey,  Pauling, and  Koltun \cite{corey1953molecular}.  Solvent accessible surface (SAS) and solvent excluded surface were proposed \cite{lee1971interpretation, richards1977areas} to provide a more elaborate 3D description of biomolecular structures. However, these surface definitions admit geometric singularities, which lead to computational instability. Smooth surfaces, including Gaussian surface \cite{Blinn1982AGO, Duncan1993ShapeAO, zheng2012biomolecular,MXChen:2012}, skinning surface \cite{cheng2009quality}, minimal molecular surface \cite{Bates:2008} and  flexibility-rigidity index (FRI)  surface \cite{KLXia:2013d,DDNguyen:2016b}, were constructed to mitigate the computational difficulty.

Another important property of macromolecules is their structural instability or flexibility. Such property measures macromolecular intrinsic ability to respond to external stimuli. Flexibility is known to be crucial for biomolecular binding, reactivity, allosteric signaling, and order-disorder transition \cite{JMa:2005}. It is typically studied by standard techniques, such as  normal mode analysis  (NMA)
 \cite{Go:1983,Tasumi:1982,Brooks:1983,Levitt:1985},   Gaussian network model (GNM) \cite{Bahar:1997} and  anisotropic network model (ANM) \cite{Atilgan:2001}. These methods have the computational complexity of ${\mathcal O}(N^3)$, with $N$ being the number of unknowns.  As a geometric graph-based method, FRI was introduced to reduce the computational complexity and improve the accuracy of GNM \cite{KLXia:2013d,  Opron:2014}. Nonetheless, NMA and ANM offer the collective motions which as manifested in normal modes, may facilitate the functionally important conformational variations of macromolecules.

The aforementioned Gaussian surface or FRI surface provides us a manifold structure embedded in 3D, which makes the analysis of geometry and topology accessible by differential geometry and algebraic topology. Recently, differential geometry has introduced to understand macromolecular structure and function  \cite{XFeng:2012a, KLXia:2014a}. In general, protein surface has many atomic scale concave and convex regions which can be easily characterized by Gaussian curvature and/or mean curvature. In particular,  the concave regions of a protein surface at  the scale of a few residue are potential ligand binding pockets.  Differential geometry-based algorithms in both Lagrangian and Cartesian formulations  have been developed to generate multiscale representations of biomolecules.   Recently,  a geometric flow based algorithm has been proposed to detect protein binding pockets \cite{Zhao2018ProteinPD}.  Morse function and Reeb graph are employed to characterize the hierarchical pocket and sub-pocket structure  \cite{Zhao2018ProteinPD,DFW13b}.

More recently, persistent homology \cite{CZCG05,ComputationalTopologyBook}, a new branch of algebraic topology,  has become a popular approach for the topological simplification of macromolecular structural complexity \cite{yao2009topological, KLXia:2014c,KLXia:2015a}. Topological invariants are macromolecular connected components, rings, and cavities. Topological analysis is able to unveil the topology-function relationship, such as ion channel open/ close, ligand binding/disassociation, and protein folding/unfolding.
However, persistent homology neglects chemical and biological information during its geometric abstraction. Element-specific persistent homology has been introduced to retain crucial chemical and biological information during the topological simplification \cite{ZXCang:2017b}. It has been integrated with deep learning to predict various biomolecular properties, including protein-ligand binding affinities and protein folding stability changes upon mutation \cite{ZXCang:2017c}.

It is interesting to note that most current theoretical models for macromolecules are built from classical mechanics,  namely, computational electromagnetics, fluid mechanics,  elasticity theory, and molecular mechanics based on Newton's law. These approaches lead to multivalued scalar, vector and tensor fields, such as macromolecular electrostatic potential, ion channel flow, protein anisotropic motion, and molecular dynamics trajectories. Biomolecular cryogenic electron microscopy (cryo-EM) maps are also scalar fields.  Mathematically, macromolecular multivalued scalar, vector, and tensor fields contain rich geometric, topological, stability, flexibility and natural mode information that can be analyzed to reveal molecular function. 
Unfortunately, unified geometric and topological analysis of macromolecular multivalued fields remains scarce. It is more challenging to establish a unified mathematical framework to further analyze macromolecular flexibility and natural modes.  There is a pressing need to develop a unified theory for analyzing the geometry, topology, flexibility, and collective motion of macromolecules so that many existing methods can be calibrated to better  uncover macromolecular function, dynamics, and transport.

The objective of the present work is to construct a unified theoretical paradigm for analyzing the geometry, topology, flexibility and natural mode of macromolecules so to reveal their function, dynamics, and transport. To this end, we introduce de Rham-Hodge theory for the modeling and analysis of macromolecules. de Rham-Hodge theory is a cornerstone of contemporary differential geometry, algebraic topology,  geometric algebra, and spectral geometry. It provides not only the Helmholtz-Hodge decomposition to uncover the interplay between geometry and topology and the conservation of certain physical observables, but also the spectral representation of the underlying multivalued fields, further unveiling the geometry and topology. Specifically, as a ubiquitous computational tool,  the Helmholtz-Hodge decomposition of various vector fields, such as electromagnetic fields \cite{hekstra2016electric}, velocity fields \cite{de1977hydrodynamic}, and deformation fields \cite{atilgan2001anisotropy}, can reveal their underlying geometric and topological features  (see a survey \cite{bhatia2013helmholtz}). Additionally,  de Rham-Hodge theory interconnects classic differential geometry, algebraic topology and partial differential equation (PDE) and provides a high-level representation of vector calculus and conservation law in physics. Finally, the spectra of Laplace-ed Rham operators in various differential forms also contain the underlying geometric and topological information and provides a starting point for the theoretical modeling of macromolecular flexibility and natural modes.  The corresponding computational tool is  discrete exterior calculus (DEC) \cite{hirani2003discrete, desbrun2005discrete, arnold2006finite}. de Rham-Hodge theory has had great success in theoretical physics, such as electrodynamics, gauge theory, quantum field theory, quantum gravity, etc. However, this versatile mathematical tool has not been applied to biological macromolecules, to the best of our knowledge.  The proposed de Rham-Hodge framework seamlessly unifies previously developed differential geometry, algebraic topology,  spectral graph theory, and PDE based approaches for biological macromolecules \cite{xia2016review}.

 Our specific contributions are summarized as follows
\begin{itemize}
    \item We provide a spectral analysis tool based on de Rham-Hodge theory to extract geometric and topological features of macromolecules. In addition to the traditional spectra of scalar Hodge Laplacians, we enrich the spectra by using vector Hodge Laplacians with various boundary conditions.
    
    \item We construct a  de Rham-Hodge theory-based analysis tool for the orthogonal decomposition of various vector fields, such as electric field, magnetic field, velocity field from molecular dynamics and displacement field,  associated with macromolecular modeling, analysis, and computation.
    
    \item We propose a novel multiscale flexibility model based on the spectra of various Laplace-de Rham operators. This new method is applied to the Debye-Waller factor prediction of a set of 364 proteins  \cite{Opron:2014}. By comparison with experimental data, we show that our new model outperforms GNM, the standard bearer in the field \cite{Bahar:1997, Opron:2014}.
    
        \item We introduce a multiscale natural mode model by constraining a vector Laplace-de Rham operator with a Helfrich curvature potential. The resulting  Laplace-de Rham-Helfrich operator is applied to analyzing the natural modes of cryo-EM data. Unlike previous normal mode analysis which assumes harmonic potential around the equilibrium, our approach allows unharmonic motions far from the equilibrium.   The multi-resolution nature of the present method makes it a desirable tool for the multiscale analysis of macromolecules, protein complexes, subcellular structures, and cellular motions.
		
\end{itemize}

\section{Results}\label{sec:results}
Our results are twofold: we first describe our contribution to computational tools for Place-de Rham operators based on the simplicial tessellation of volumes bounded by biomolecular surfaces, then we present
the modeling and analysis of de Rham Hodge theory for biological macromolecules.

\subsection{Theoretical modeling and analysis}

This section introduces de Rham-Hodge  theory for the analysis of biomolecules. To establish notation, we provide a brief review of de Rham-Hodge theory. Then, we introduce topological structure-preserving analysis tools, such as discrete exterior calculus (DEC) \cite{desbrun2005discrete}, discretized differential forms, and Hodge-Laplacians, on the compact manifolds enclosing biomolecular boundaries. We use simple finite-dimensional linear algebra to computationally realize our structure-preserving analysis on various differential forms.  We construct appropriate physically-relevant boundary conditions on biomolecular manifolds to facilitate various scalar and vector Laplace-de Rham operators such that the resulting spectral bases are consistent with three basic singular value decompositions of the gradient, curl and divergence operators through dualities.

\paragraph{de Rham-Hodge theory for macromolecules}
While the spectral analysis can be carried out using scalar, vector and tensor calculus, differential forms and exterior calculus are required in de Rham-Hodge theory to reveal the intrinsic relations between differential geometry and algebraic topology on biomolecular manifolds. Since biomolecular shapes can be described as compact manifolds in the 3D Euclidean space, we represent scalar and vector fields on molecular manifolds as well as their interconversion through differential forms. As a generalization to line integral and flux calculation of vector fields, a differential $k$-form $\omega^k\in\Omega^k(M)$ is a field that can be integrated on a $k$-dimensional submanifold of $M$, which can be mathematically defined through a rank-$k$ antisymmetric tensor defined on a manifold $M$. By treating it as a multi-linear map from $k$ vectors spanning the tangent space to a scalar, it turns an infinitesimal $k$-dimensional cell into a scalar, whose sum over all cells in a tessellation of a $k$-dimensional submanifold produces the integral in the limit of infinitesimal cell size. In ${\mathbb R}^3$, $0$-forms and $3$-forms have one degree of freedom at each point and can be regarded as scalar fields, while $1$-forms and $2$-forms have three degrees of freedom, and can be interpreted as vector fields.

The differential operator (also called exterior derivative) $d$ can be seen as a unified operator that corresponds to gradient ($\nabla$), curl ($\nabla\!\times$) , and divergence ($\nabla\cdot$) when applied to $0$-, $1$-, and $2$-forms, mapping them to $1$-, $2$-, and $3$-forms, respectively. On a boundaryless manifold, a codifferential operator $\delta$ is the adjoint operator under $L_2$-inner product of the fields (integral of pointwise inner product over the whole manifold), which corresponds to $-\nabla\cdot$, $\nabla\!\times$, and $-\nabla$, for $1$-, $2$-, and $3$-forms, mapping them to $0$-, $1$-, and $2$-forms, respectively.

One key property of $d:\Omega^k\to\Omega^{k+1}$ is that $d  d=0$, which allows the space of differential forms $\Omega^k$ to form a chain complex, which is called the de Rham complex.
\begin{equation}
	0 \xrightarrow{}  \Omega^{0}(M) \xrightarrow[(\nabla)]{d} \Omega^{1}(M) \xrightarrow[(\nabla \times)]{d} \Omega^{2}(M) \xrightarrow[(\nabla \cdot)]{d} \Omega^{3}(M) \xrightarrow{d} 0.
\end{equation}

It also matches the identities of second derivatives for vector calculus in ${\mathbb R}^3$, i.e., $(\nabla \times) \nabla = 0$ and $(\nabla \cdot) \nabla \times = 0$.  The topological property associated differential forms is given by the de Rham cohomology,
$$H_{dR}^k(M)=\frac{\ker d^k}{\textrm{im} d^{k-1}}.$$
The de Rham theorem states that the de Rham cohomology is isomorphic to the singular cohomology, which is derived purely from the topology of the biomolecular manifold. Hodge further established the isomorphism
$$H_{dR}^k(M)\cong H_\Delta^k(M),$$
where $H_\Delta^k(M)=\{\omega|\Delta \omega = 0\}$ is the kernel of the Laplace-de Rham operator $\Delta \equiv d\delta+\delta d=(d  + \delta )^2$, also known as the space of  harmonic forms. Here,
$\delta:\Omega^k\to\Omega^{k-1}$ satisfies $\delta   \delta=0$.
A corollary of the result is the Hodge decomposition,
\begin{equation}
\omega=d\alpha+\delta\beta + h,
\end{equation}
which is an $L_2$-orthogonal decomposition of any form $\omega$ into $d$ and $\delta$ of two potential fields $\alpha \in\Omega^{(k-1)} (M)$ and $\beta\in\Omega^{(k+1)} (M)$ respectively, and a harmonic form $h\in H_\Delta^k(M)$. This means that harmonic forms are the non-integrable parts of differential forms, which form a finite dimensional space determined by the topology of the biomolecular domain due to de Rham's and Hodge's theorems.

\begin{figure*}
	\centering
	\includegraphics[width=.98\linewidth]{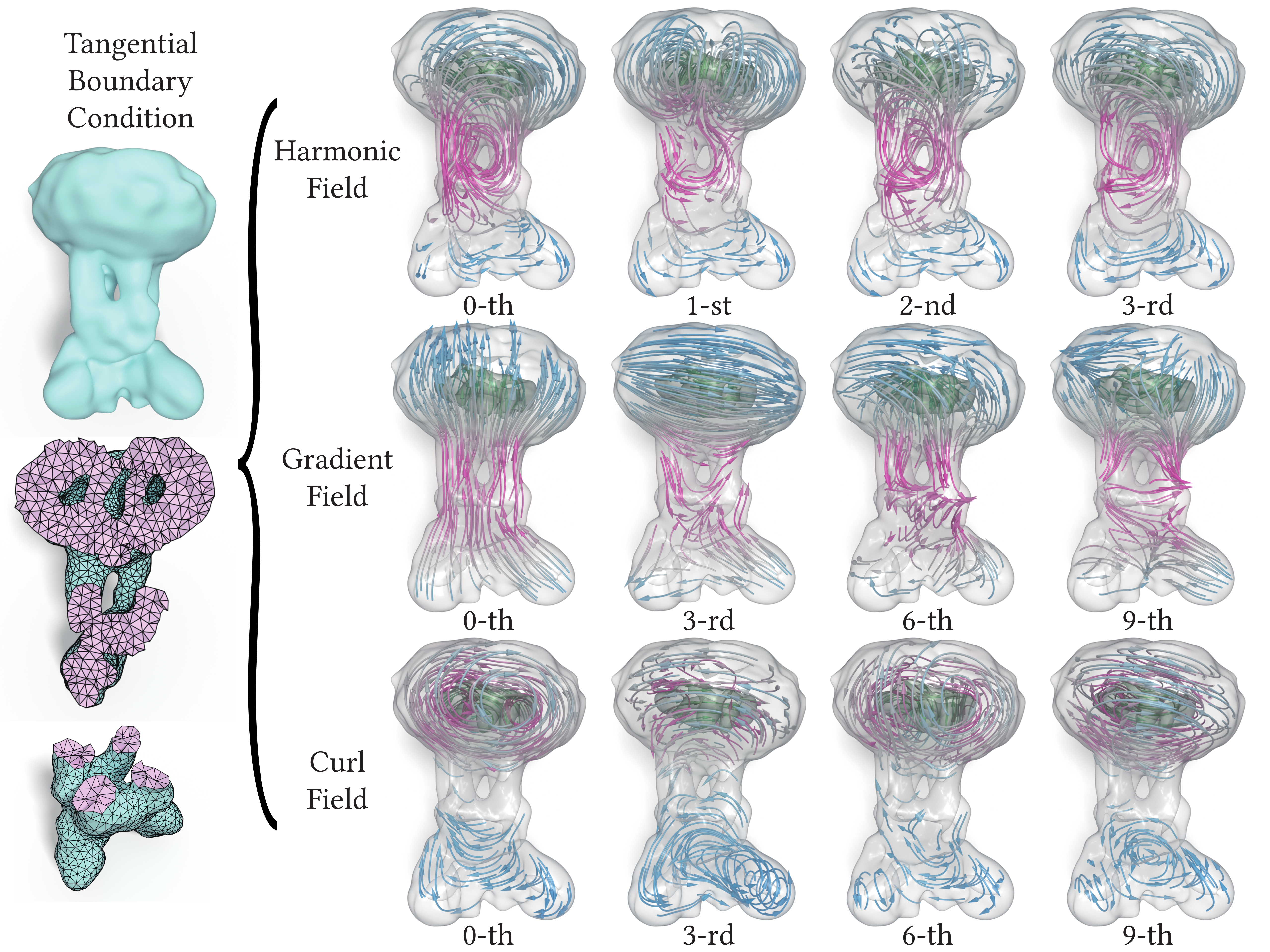}
	\caption{\textbf{Illustration of tangential spectra of a cryo-EM map EMD 7972} Topologically, EMD 7972  has 6 handles and 2 cavities. The left column is the original shape and its anatomy showing the topological complexity. On the right-hand side of the parenthesis, the first row shows tangential harmonic eigen fields, the second row shows tangential gradient eigen fields, and the third row shows tangential curl eigen fields.}
	\label{fig:L1t}
\end{figure*}

\paragraph{Macromolecular spectral analysis}

The Laplace-de Rham operator $\Delta=d\delta+\delta d$, when restricted to a 3D object embedded in the 3D Euclidean space, is simply $-\nabla^2$. As it is a self-adjoint operator with a finite dimensional kernel, it can be used to build spectral bases for differential forms. For irregularly shaped objects, these bases can be very complicated. However, for a simple geometry, these bases are well-known functions. For example, $0$-forms on a unit circle can be expressed as the linear combination of sine and cosine functions, which are eigenfucntions of the Laplacian for $0$-forms $\Delta_0$. Similarly, spherical harmonics are eigenfunctions of $\Delta_0$ on a sphere and it has also been extended  to manifold harmonics on Riemannian  2-manifolds.

We further extend the analysis to any rank $k$ and to 3D shapes such as macromolecular shapes where analysis can be carried out in two types of cases. In the first type, one may treat the surface of the molecular shape as a boundaryless compact manifold and analyzes any field defined on such a 2D surface. In fact, this approach is relevant to protein surface electrostatic potentials or the behavior of cell membrane or mitochondrial ultrastructure. In this work, we shall restrain from any further exploration in this direction. In the second type, we consider the volumetric data enclosed by a macromolecular surface. As a result, the molecular shape has a boundary. In this setting, the harmonic space becomes infinite-dimensional unless certain boundary conditions are enforced. In particular, tangential or normal boundary conditions (also called  Dirichlet or  Neumann boundary conditions, respectively) are enforced to turn the harmonic space into a finite-dimensional space corresponding to algebraic topology constructions that lead to absolute and relative homologies.

We first discuss the natural separation of the eigenbasis fields into curl-free and div-free fields in the continuous theory, assuming that the boundary condition is implicitly enforced, before providing details on the discrete exterior calculus with the boundary taken into consideration.

Given any eigen field $\omega$ of the Laplacian,
$$\Delta \omega = \lambda \omega,$$
we can decompose it into $\omega=d\alpha+\delta\beta+h$. For $\lambda\neq 0$, $h=0$, based on $dd=0$ and $\delta\delta=0$, it is easy to see that
both $d\alpha$ and $\delta\beta$ are eigenfunctions of $\Delta$ with eigenvalue $\lambda$ due to the uniqueness of the decomposition, unless one of them is $0$. It is typically the case that $\omega$ is either a curl field or a gradient field, otherwise, $\lambda$ has a multiplicity of at least $2$, in which case both eigen fields associated with $\lambda$ are the linear combinations of the same pair of the gradient field and the curl field.

\paragraph{Discrete spectral analysis of differential forms}\label{sec:DiscreteSpectralAnalysis}

In a simplicial tessellation of a manifold mesh, $d^k$ is implemented as a matrix $D_k$, which is a signed incidence matrix between $(k\!+\!1)$-simplices and $k$-simplices. We provide the details in Sec.~\ref{Sec:Methods}, but the defining property in de Rham-Hodge theory is preserved through such a discretization: $D_{k+1}D_k=0$. The adjoint operator $\delta^k$ is implemented as $S_{k-1}^{-1}D_{k-1}^T S_k$, where $S_k$ is a mapping from a discrete $k$-form to a discrete $(n-k)$-form on the dual mesh, which can be treated as a discretization of the $L_2$-inner product of $k$-forms. As $S_k$ is   always a symmetric positive matrix, the $L_2$-inner product between two discrete $k$-forms can be expressed as $(\omega^k_1)^T S_k \omega^k_2$.  The discrete Hodge Laplacian is defined as
$$L_k=D_k^T S_{k+1} D_k + S_k D_{k-1} S_{k-1}^{-1} D_{k-1}^T S_k,$$
which is a symmetric matrix and $S_k^{-1} L_k$ corresponds to $\Delta_k$. The eigenbasis functions are found through a generalized eigenvalue problem,
\begin{equation}\label{eqn:GEP}
L_k \omega^k = \lambda^k S_k \omega^k.
\end{equation}
Depending on whether the tangential or normal boundary condition is enforced, $D_k$ includes or excludes the boundary elements respectively. Thus, the boundary condition is built into discrete linear operators. When we need to distinguish these two cases, we use $L_{k,t}$ and $L_{k,n}$ to denote the tangential and normal boundary conditions respectively.

In general, it is not necessarily efficient to take the square root of a symmetric matrix $S_k$ or to compute its inverse. However, for analysis, we can always convert a generalized eigenvalue problem in Eq. (\ref{eqn:GEP}) into a regular eigenvalue problem,
$$\bar{L}_k \bar{\omega}^k\equiv S_k^{-\frac12} L_k S_k^{-\frac12} \bar{\omega}^k = \lambda^k \bar{\omega}^k,$$
where $\bar{\omega}\equiv S_k^{\frac12}\omega$. We can further rewrite the symmetrically modified Hodge Laplacian as
\begin{equation}\label{eqn:HLaplacian}
\bar{L}_k = \bar{D}_k^T \bar{D}_k+ \bar{D}_{k-1} \bar{D}_{k-1}^T,
\end{equation}
where $\bar{D}_k \equiv S_{k+1}^{\frac12} D_k S_{k}^{-\frac12}$ must satisfy $\bar{D}_{k+1}\bar{D}_k = 0$. Now the $L_2$-inner product between two discrete differential forms in the modified space is simply $(\omega^k_1)^T \omega^k_2$, and the adjoint operator of $\bar{D}_k$ is simply $\bar{D}_k^T$.

Now the partitioning of the eigenbasis fields into harmonic fields, gradient fields, and curl fields for $1$-forms and $2$-forms and their relationship can be understood from the singular value decomposition of the differential operator
$$\bar{D}_k =U_{k+1}\Sigma_k V_k^T,$$
where $U_{k+1}$ and $V_k$ are orthogonal matrices, and $\Sigma_k$ is a rectangular matrix that only has nonzero entries on the diagonal, which can be sorted in ascending order as $ \sqrt{\lambda^k_{i}}$ with trailing zeros. As the Hodge decomposition is an orthogonal decomposition, each column of $V_k$ that corresponds to a nonzero singular value $\sqrt{ \lambda^k_{i}}$ is orthogonal to any column of $U_k$ that corresponds to a nonzero $\sqrt{\lambda^{k-1}_j}.$  Here $V_k$ and $U_k$,  together with the finite dimensional set of harmonic forms $h_k$ (which satisfy both $D_k h_k=0$ and $D_{k-1}^T h_k=0$), span the entire space of $k$-forms. Moreover, the spectrum (i.e., set of eigenvalues) of
the symmetric modified Hodge Laplacian in Eq. (\ref{eqn:HLaplacian})
consists of $0$s,  the set of $\lambda^k_{i}$'s, and the set of $\lambda^{k-1}_j$'s.
Note that, in the spectral basis, taking derivatives $\bar{D}$ (or $\bar{D}^T$) is simply performed through multiplying the corresponding singular values, and integration is done through division by the corresponding singular values, mimicking the situation in the traditional Fourier analysis for scalar fields.

\begin{figure*}
	\centering
	\includegraphics[width=.98\linewidth]{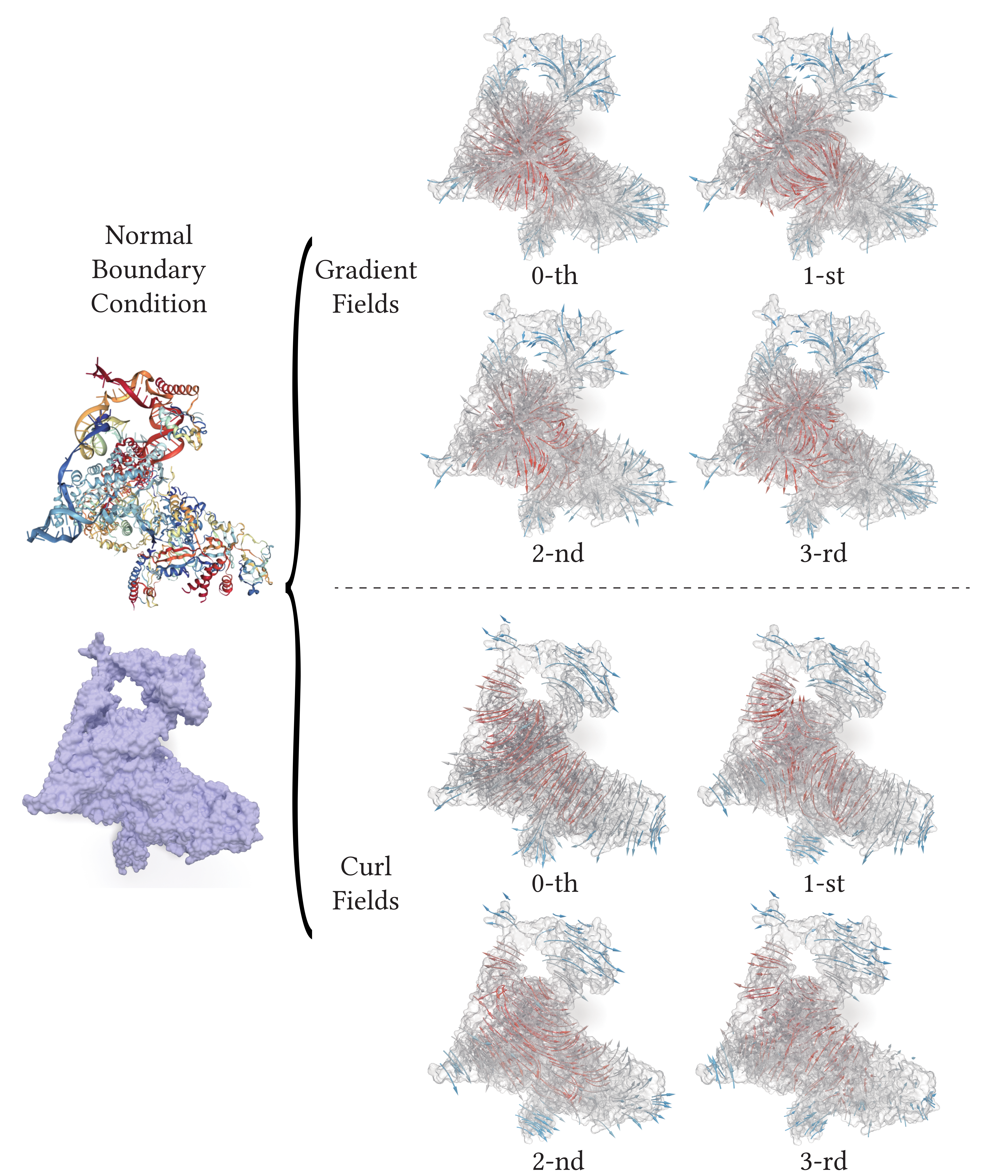}
	\caption{\textbf{Illustration of the normal spectra of protein and DNA complex 6D6V} Topologically,    the crystal structure of 6D6V  has 1 handle. The left column shows the secondary structure and the solvent excluded surface (SES). On the right-hand side, the first two rows show normal gradient eigen fields, and the last two rows show normal curl eigen fields.}
	\label{fig:L1n}
\end{figure*}

\paragraph{Boundary conditions and dualities in 3D molecular manifolds}\label{sec:BoundaryDuality}

Overall, appropriate boundary conditions are prescribed to preserve the orthogonal property of the Hodge decomposition. In 3D molecular manifolds, for the spectral analysis of scalar fields (0-forms or 3-forms), two types of typical boundary conditions are used: Dirichlet boundary condition $f|_{\partial M}=f_0$ and Neumann boundary condition $\bn\cdot\nabla f|_{\partial M}=g_0$, where $f_0$ and $g_0$ are functions on the boundary $\partial M$ and $\bn$ is the unit normal on the boundary. For spectral analysis, harmonic fields satisfying the arbitrary boundary conditions can be dealt with through spectral analysis of $f_0$ or $g_0$ on the boundary, and the following boundary conditions are used for the volumetric function $f$. The normal 0-forms (tangential 3-forms) satisfy
$$f|_{\partial M}=0,$$
and the tangential 0-forms (normal 3-forms) satisfy
$$\bn\cdot\nabla f|_{\partial M}=0.$$

For the spectral analysis of vector fields, boundary conditions are for the three components of the field.  Based on the de Rham-Hodge theory, it is more convenient to also use two types of boundary conditions. For tangential 1-forms (normal 2-forms) $\bv$, we use the Dirichlet boundary condition for the normal component and the Neumann condition for the tangential components:
$$\bv\!\cdot\!\bn\!=\!0,\;\; \bn \!\cdot\! \nabla (\bt_1 \cdot \bv) \!=\! 0,\;\; \bn \!\cdot\! \nabla (\bt_2 \cdot \bv)=0,$$
where $\bt_1$ and $\bt_2$ are two local tangent directions forming a coordinate frame with the unit normal $\bn$. The corresponding spectral fields are shown in Fig \ref{fig:L1t}. For normal 1-forms (tangential 2-forms) $\bv$, we use the Neumann boundary condition on the normal component, and the Dirichlet boundary condition on the tangential components:
$$\bv \!\cdot\! \bt_1\!=\!0,\;\; \bv \!\cdot\! \bt_2\!=\!0,\;\;\bn\!\cdot\!\nabla\bv_{\bn}\!=\!0.$$
The corresponding spectral fields are shown in Fig. \ref{fig:L1n}. Despite of the harmonic spectral  fields, there are two types of fields involved for the spectral fields of both boundary conditions. One is the set of divergence-free fields and the other is the set of curl-free fields.

\begin{figure*}
	\centering
	\includegraphics[width=.98\linewidth]{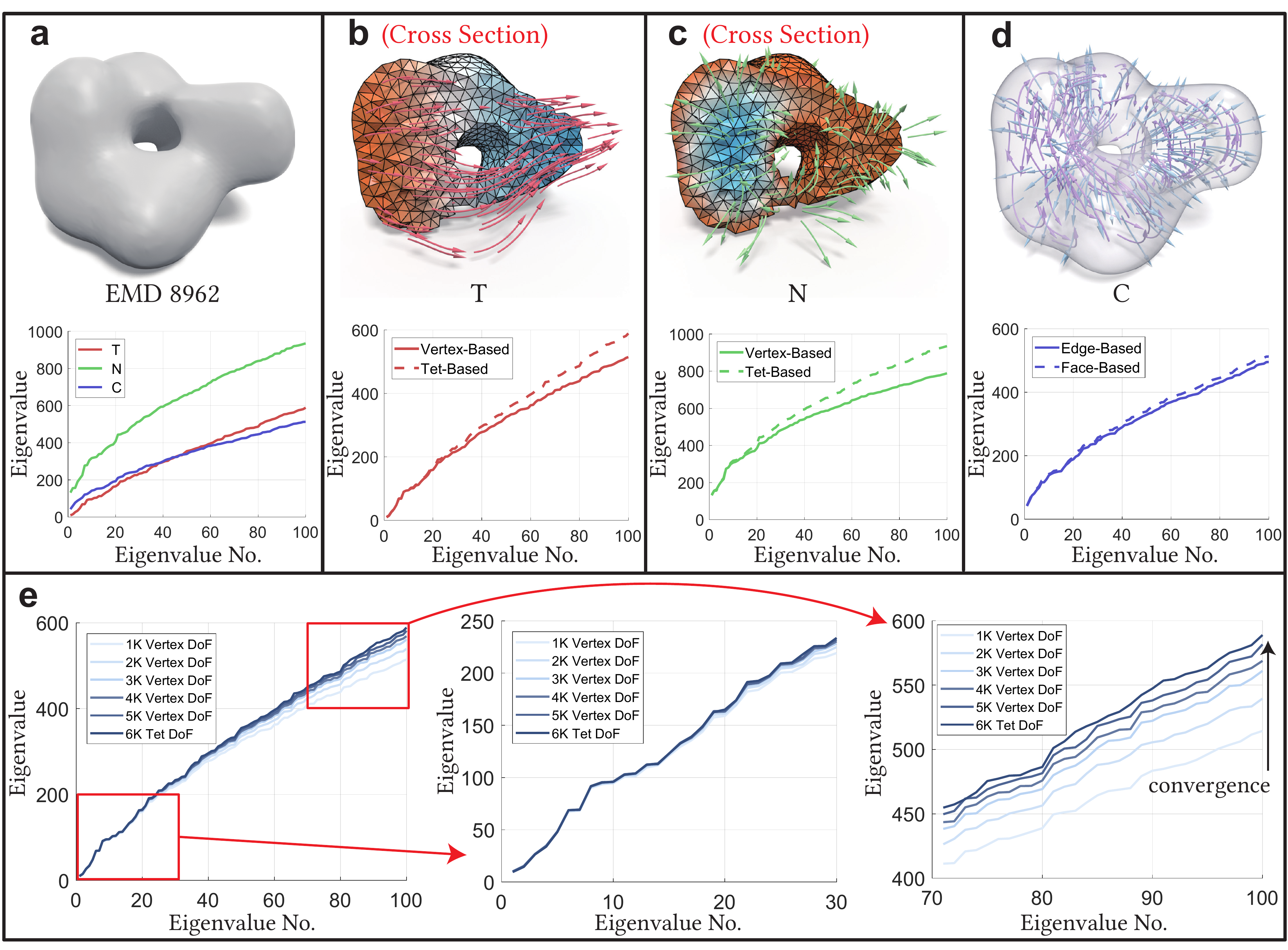}
	\caption{\textbf{Illustration of Hodge Laplacian spectra} This figure shows the properties of 3 spectral groups, namely, tangential gradient eigen fields ($T$), normal gradient eigen fields ($N$), and curl eigen fields ($C$), for EMD 8962.
\textbf{a} shows the original input surface and 3 distinct spectral groups.
\textbf{b} shows the cross section of a typical tangential gradient eigen field and the distribution of eigenvalues for group $T$.
\textbf{c} shows the cross section of a typical normal gradient eigen field and the distribution of eigenvalues for group $N$.
\textbf{d} shows   a typical curl eigen field and the distribution of eigenvalues for group $C$.
\textbf{e} The left chart shows the convergence of spectra in the same spectral group due to the increase in the mesh size, i.e., the DoFs  from 1,000 (1K) to 6,000 (6K).  Obviously, low order eigen modes converge fast (middle chart) and high order eigen modes converge slowly (right chart).
 }
	\label{fig:spectrum}
\end{figure*}

\paragraph{Reduction and analysis}
For the four types of $k$-forms ($k\in\{0,1,2,3\}$ in ${\mathbb R}^3$) in combination with the two types of boundary conditions (tangential and normal), there are 8 different Laplace-de Rham operators ($L_{k,t}$ and $L_{k,n}$) in total.  However, based on the discussion in \ref{sec:DiscreteSpectralAnalysis}, the nonzero-parts of the spectrum $L_{k}$ can be assembled from the singular values of $\bar{D}_k$ and $\bar{D}_{k-1}$ (See \ref{eqn:HLaplacian}). Thus, for each type of boundary condition, there are only three spectra associated with $\bar{D}_0$, $\bar{D}_1$, and $\bar{D}_2$, since $\bar{D}_3=0$ for 3D space (However, one still has eight $L$ operators).  Moreover, according to the duality discussed in \ref{sec:BoundaryDuality}, there is a one-to-one mapping between tangential $k$-forms and normal $(3\!-\!k)$-forms, which further identifies $\bar{D}_{0,t}$ with $\bar{D}^T_{2,n}$, $\bar{D}_{0,n}$ with $\bar{D}^T_{2,t}$, and $\bar{D}_{1,t}$ with $\bar{D}^T_{1,n}$. As a result, one has four independent $L$ operators. Finally, due to the self-adjointness, there are only 3 intrinsically different spectra: 1) The first contains singular values of the gradient operator on tangential scalar potential fields   (or equivalently, the singular values of the divergence operator on tangential gradient fields) as shown in \ref{fig:spectrum}~b; 2) The second contains singular values of the gradient operator on normal scalar potential fields   (or equivalently, the singular values of the divergence operator on normal gradient fields) as shown in \ref{fig:spectrum}~c; The third contains singular values of the curl operator applied to tangential curl fields  (or equivalently, the singular values of the curl operator applied to normal curl fields) as shown in \ref{fig:spectrum}~d.

As the discussed above, each of the 8 Hodge Laplacians defined for smooth fields on a smooth shape has a spectrum that is simply the combination of one or two of the 3 sets of singular values along with possibly a $0$. However, the numerical evaluation of the singular values of the differential operators for tangential $k$-forms $\bar{D}_{k,t}$ can differ from those of the discrete operators for normal $3-k$-forms $\bar{D}^T_{2-k,n}$, as shown in \ref{fig:spectrum}~d. One immediate reason is that the degrees of  freedom  (DoFs) associated for tangential/normal scalar/vector fields represented as tangential forms are not the same as those represented by normal forms on a given tessellation, leading to different sampling accuracies. For example, the tessellation of the shape in (Fig. \ref{fig:spectrum}) consists of approximately $1,000$ vertices, $7,000$ edges, $10,000$ triangles and $5,000$ tetrahedra. Thus, each tangential $0$-form only has $1,000$ DoFs, and each normal $3$-form has $5,000$. Hence, $L_{3,n}$ is capable of handling higher frequency signals in any given smooth scalar field than $L_{0,t}$ when we approach the Nyquist frequencies of the sampling. The convergence of both discretizations for the same continuous operator can be observed with increasing DoFs for both differential forms under refinement of the tet meshes (\ref{fig:spectrum}~e left). For low frequencies, there is a good agreement to begin with (\ref{fig:spectrum}~e mid), while for any given high frequency, the convergence with increased resolutions can be clearly observed (\ref{fig:spectrum}~e right).

On the other hand, $\bar{D}_{k}\bar{D}^T_{k}$ and $\bar{D}_{k}\bar{D}^T_{k}$ will have strictly the same set of nonzero eigenvalues. For instance,  the spectrum of $L_{0,t}$ and the partial spectrum of $L_{1,t}$ that corresponds to gradient fields are identical, since $\bar{D}_{0,t}\bar{D}_{0,t}^T$ and $\bar{D}_{0,t}^T\bar{D}_{0,t}$ have the same nonzero eigenvalues.

For eigen fields vector Laplacians represented as $1$-forms or $2$-forms, i.e. the eigen fields of $L_1$ or $L_2$, we can observe some typical traits in the distributions of eigenvalues under normal and tangential boundary conditions. The normal boundary condition tends to allow more gradient eigen fields associated with eigenvalues below a given threshold than those under the tangential boundary condition for eigenvalues below the same threshold. We conjecture that it is due to the more stringent Dirichlet boundary condition on the potential scalar fields than the Neumann boundary condition on the potential scalar fields. The relation between the tangential boundary condition gradient-type eigen fields and curl-type eigen fields for low-frequency range seems to be highly dependent on the shape (Fig. \ref{fig:spectrum}~f). Fig. \ref{fig:L1t} shows different vector eigen fields for tangential boundary condition with EMD 7972 surface. The first row shows different harmonic fields corresponding to the number of handles of the shape, the second row shows different gradient fields and the third row shows different curl fields. Fig. \ref{fig:L1n} shows different vector eigen fields for normal boundary condition with the protein and DNA complex crystal structure 6d6v. Since there are no cavities for this shape, there are no harmonic fields. The first row shows different gradient fields and the second row shows different curl fields. Note that the scalar potentials for gradient fields and the vector potentials for curl fields are also themselves eigen fields associated with the same eigenvalues, although for different Laplacians.

Summarizing the above discussion on the properties of Laplacian spectra for 3D shape, we propose the following suggestions for practical spectral analysis.
\begin{itemize}
	\item Only 3 independent spectra (e.g., singular value spectra of $D_{0,t}$, $D_{1,t}$, and $D_{2,t}$) are necessary to avoid redundancy.
	\item Laplace-de Rham operators with higher DoFs can be used for more accurate calculation (at a higher computational cost) given the same tessellation.
	\item When computing eigenvalues given the same high-frequency truncation threshold, the differences in the numbers of eigenvalues in the 3 spectra vary with the shape.
\end{itemize}

\subsection{Macromolecular modeling and analysis}

Biological macromolecules and their complexes offer a rich variety of geometric and topological features, which often exhibit close relations with their functionalities. For instance, protein pockets can often be identified as a geometrically concave region on the protein surface, or as a topological cavity of an offset surface. Ion channels that regulate important biological functions can be usually associated with a topological tunnel. Mitochondrial ultrastructures admit various geometric and topological complexity  which is related to their functions. Hence, a unified approach for quantitatively analyzing such geometric and topological features is in great need. Our de Rham-Hodge analysis and Laplace-de Rham operator modeling provide such a unified approach for capturing both geometric and topological features simultaneously.

Our de Rham-Hodge analysis offers a powerful new tool for characterizing macromolecular geometry, identifying macromolecular topology, and modeling macromolecular structural flexibility and collective motion. We have carried out extensive computational experiments using protein structural datasets and cryo-EM maps to demonstrate the utility and usefulness of the proposed de Rham-Hodge tools and models.

\paragraph{Molecular shape generation}

The geometric modeling of macromolecular 3D  shapes bridges the gap between experimental data and theoretical models for macromolecular function, dynamics, and transport.  To carry out our de Rham-Hodge analysis on a macromolecule or a protein complex, we need a given domain containing the 3D macromolecular shape. Theoretically, such a domain for a macromolecule can be generated by taking the isosurface of a cryo-EM map or constructed from the atomic coordinates of the macromolecule. For a given set of atomic coordinates ${\bf r}_i, i=1,2,\cdots, N$, van der Waals surface, solvent accessible surface, and the solvent excluded surface can be constructed. However, these surfaces are typically singular, leading to computational instability for de Rham-Hodge analysis. Alternatively, minimal molecular surface  (MMS) generated by differential geometry, Gaussian surface, and flexibility rigidity index (FRI) surface \cite{KLXia:2013d,Opron:2014} are computationally preferred. In fact, FRI surface is simpler than MMS and more stable than Gaussian surface \cite{DDNguyen:2016b}. To generate an FRI surface, we use a discrete-to-continuum mapping to define an unnormalized molecular density \cite{KLXia:2013d,DDNguyen:2016b}
\begin{equation}
\rho({\bf r}, \eta)=\sum_{j=1}^N\Phi(\| {\bf r}-{\bf r}_j \|;\eta)
\end{equation}
where $\eta$ is a scale parameter and is set to twice of the atomic van der Waals radius $r_j$.
$\Phi$ is  density estimator  that satisfies the following admissibility conditions
\begin{align}\label{eq:admiss}
\Phi \left(\|\mathbf{r}- \mathbf{r}_j\|;\eta \|\right)&=1, \quad{\rm as} \quad  \|\mathbf{r} -\mathbf{r}_j\| \rightarrow 0, \\
\Phi \left(\|\mathbf{r} - \mathbf{r}_j\|;\eta \|\right)&=0, \quad {\rm as} \quad  \|\mathbf{r} -\mathbf{r}_j\| \rightarrow \infty.
\end{align}
Monotonically decaying  radial basis functions are all admissible. Commonly used  correlation kernels include  generalized exponential functions
\begin{align}\label{exponential}
\Phi\left(\|\mathbf{r} -\mathbf{r}_j\|;\eta \|\right)=e^{-\left(\|\mathbf{r} -\mathbf{r}_j\|/\eta \right)^\kappa}, \quad \kappa>0;
\end{align}
and generalized Lorentz functions
\begin{align}\label{Lorentz1}
\Phi\left(\|\mathbf{r} -\mathbf{r}_j\|;\eta \right)=\frac{1}{1+\left(\|\mathbf{r} -\mathbf{r}_j\|/\eta \right)^\nu},\quad \nu>0.
\end{align}
 The Gaussian kernel ($\kappa=2$) is employed in this work.

A family of biomolecular domains can be defined by varying  level set parameter $c>0$
 \begin{equation}\label{manifold}
M= \{{\bf r}|   \rho({\bf r},  \eta)=c\}.
\end{equation}

\paragraph{Topological analysis}

\begin{figure}
	\centering
	\includegraphics[width=.98\linewidth]{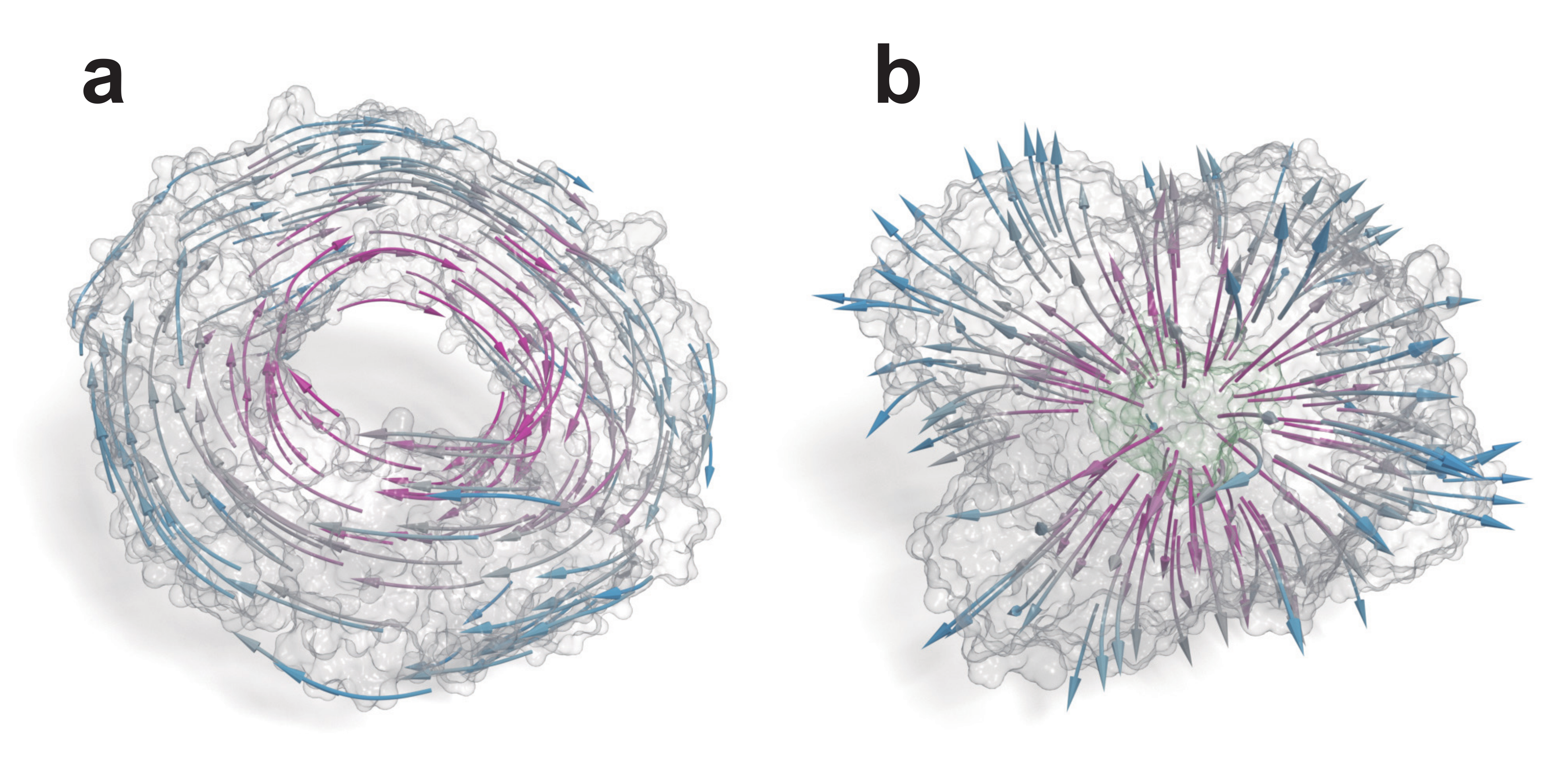}
	\caption{\textbf{Illustration of topological analysis}. \textbf{a} Eigen fields by null space of tangential Laplace-de Rham operators correspond to handles. \textbf{b} Eigen fields by null space of normal Laplace-de Rham operators correspond to cavities. }
	\label{fig:homology}
\end{figure}

  In this work, we discuss topology in the mathematical sense. Therefore, topological features are those stable structural characteristics that do not change with deformation, such as the number of connected components, the number of holes on each connected components, and the number of cavities. They are captured in the null spaces of the corresponding Laplace-de Rham operators. In other words, the invariant spaces associated with the eigenvalue of 0, i.e., the lowest ends of the spectra. Specifically, the dimension of the null space of $L_{1,t}$ and $L_{2,n}$ is the same as the number of tunnels as shown in Fig. \ref{fig:homology} \textbf{a}.  The dimension of the null space of $L_{1,n}$ and $L_{2,t}$ provides the number of cavities as shown in Fig. \ref{fig:homology} \textbf{b}.  The dimension of the $L_{0,t}$ is equal to the number of connected components. In persistent homology, the geometric measurement for characterizing  the persistence of a topological feature has been proven crucial to the practical use of these otherwise overly stable features. The eigen fields associated with the eigenvalue 0 in our spectral analysis can also provide such information, for instance, the strength of the eigen vector field associated with the eigenvalue 0 for $L_{1,t}$ can indicate how narrow the handle/tunnel is in the region.

\paragraph{Geometric analysis}

\begin{figure}
	\centering
	\includegraphics[width=.98\linewidth]{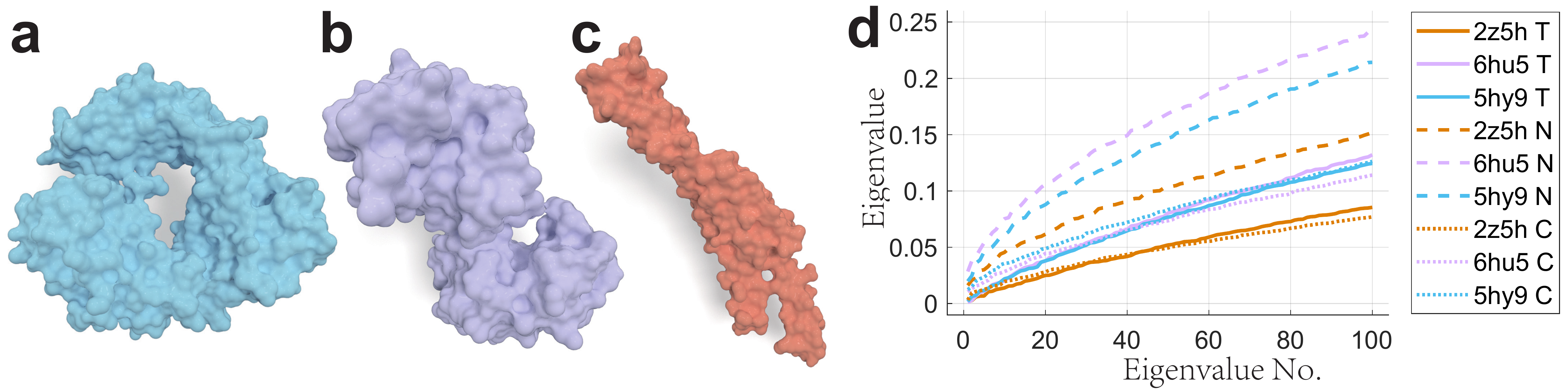}
	\caption{\textbf{Illustration of geometric analysis} The geometry of different molecules (PDB IDs: 2Z5H (\textbf{a}), 6HU5 (\textbf{b}), and 5HY9 (\textbf{c})) can be captured by three groups of different Hodge Laplacian spectra with clear separations shown in \textbf{d}. Note that the color of the line plot corresponds to the color of the molecules. The solid lines show tangential gradient (T) spectrum, the dashed lines show the normal gradient (N) spectrum, and the dot lines show the curl spectrum (C). While there is a possibility that certain spectral sets maybe close to each other (see groups T of proteins 6HU5 and 5HY9), the other 2 groups of spectra will show a clear difference. In addition, our topological features will also provide a definite difference. For example,  protein 6HU5 has trivial topology (ball), but protein 5HY9 has a handle.}
	\label{fig:geom_chara}
\end{figure}

 Although the spectra of the Laplace-de Rham operators do not uniquely determine the geometry (sometimes referred to as ``you cannot hear the shape of the drum''), they do provide key information when comparing shapes, which, sometimes, is referred to as shape ``DNA''. Thus, the traits of the non-zero parts of the spectra can be regarded as geometrical features. These geometrical features are rigid transformation invariant. The scalar Hodge Laplacian spectrum has already been used in computer graphics and computer vision to distinguish various structures in shape analysis and shape retrieval. It has also been extended to 1-form Hodge Laplacian on surfaces for the purpose of shape analysis. However, on surfaces, $L_1$ spectrum is identical to $L_0$ spectrum, except that the multiplicity is doubled for nonzero eigenvalues. Note that the multiplicity for the zero eigenvalue is determined by the number of genus instead of the number of connected components for scalar Hodge Laplacian.
In our 3D extension, we have three unique spectra for each molecule. Fig. \ref{fig:geom_chara} shows non-zero spectrum traits for 3 simple proteins (PDB IDs: 2z5h, 6hu5, and 5hy9), where the clear distinction among the spectra can be observed. We have tested on various biomolecules and observed the same discriminating ability of the spectra on these shapes.

Geometric analysis and topological analysis based on the de Rham-Hodge theory can be readily applied to characterizing biomolecules in machine learning and to biomolecular modeling. To further demonstrate the capability of de Rham-Hodge spectral analysis for  macromolecular analysis, we  propose a set of de Rham-Hodge models for  protein flexibility analysis and a vector de Rham model for biomolecular natural mode analysis.

\paragraph{Biological field decomposition and analysis}

\begin{figure*}
	\centering
	\includegraphics[width=.98\linewidth]{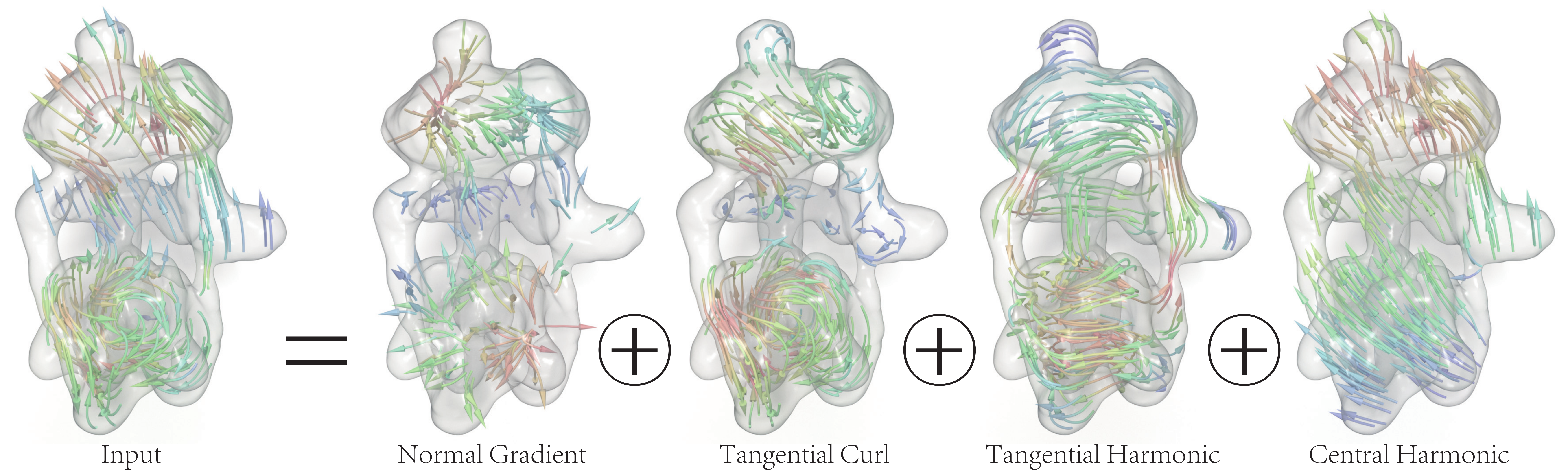}
	\caption{\textbf{Biological flow decomposition} Illustration of a synthetic vector field in EMD 1590 that is decomposed into several mutually orthogonal components based on different boundary conditions.}
	\label{fig:decom}
\end{figure*}

Our Laplace-de Rham operators constructed from different boundary conditions can also perform vector field decomposition tasks. There are naturally various vector fields existed in biomolecules, such as electric fields, magnetic fields, and elastic displacement fields. de Rham-Hodge theory can help provide a mutually orthogonal decomposition to investigate source, sink and vortex features presented in those fields. An example of this analysis is given in Fig. \ref{fig:decom} for a synthetic vector field on a vacuolar ATPase motor, EMD 1590. We expect this decomposition becomes more interesting for biomolecular electric fields,  dipolar fields, and magnetic fields. 

It is to point out that various components from the decomposition can be naturally used as the components of machine learning feature vectors.

\paragraph{Flexibility analysis}

Biomolecular flexibility analysis and B factor prediction have been commonly performed by normal mode analysis \cite{Go:1983,Tasumi:1982,Brooks:1983,Levitt:1985,JMa:2005} and Gaussian network model (GNM) \cite{Bahar:1997}.  Recently, graph theory-based FRI  has been shown to outperform other methods \cite{Opron:2014}. However, all of the aforementioned methods are based on the discrete coordinate representation of biomolecules.  As such, it is not very convenient to use these methods for flexibility analysis at different scales. For example, for some large macromolecules, such as an HIV viral capsid which involves millions of atoms, one may wish to analyze their flexibility at atomic, residue, protein domain, protein, and protein complex scales by using a unified approach so that the results from cross-scales can be compared on an equal footing. However, current approaches cannot provide such a unified cross-scale flexibility analysis. In this work, we introduce a de Rham-Hodge theory-based model to quantitatively analyze macromolecular flexibility cross many scales.

We assume that the de Rham-Hodge    B factor at the $i$th atom estimated by the $\bar{L}_k$ is given by
\begin{equation}\label{eqn:Bfactor}
B^{\rm dRH}_{k,i} = a \sum_{j }\frac{1}{\lambda^k_j}\left[\omega^k_j({\bf r}) (\omega^k_j({\bf r}'))^T   \right]_{{\bf r}={\bf r}_i , {\bf r}'={\bf r}_i }, \forall \lambda^k_j>0,
\end{equation}
where $a$ is a parameter to be determined by the least squares regression. In the computation, the value of $\omega^k_j({\bf r})$ is given on a set of mesh points. The linear regression over  a cutoff radius  $d$  is used to obtain the required values in atomic centers ${\bf r}_i$ where the B-factor values are reported.

\begin{figure*}
	\centering
	\includegraphics[width=.98\linewidth]{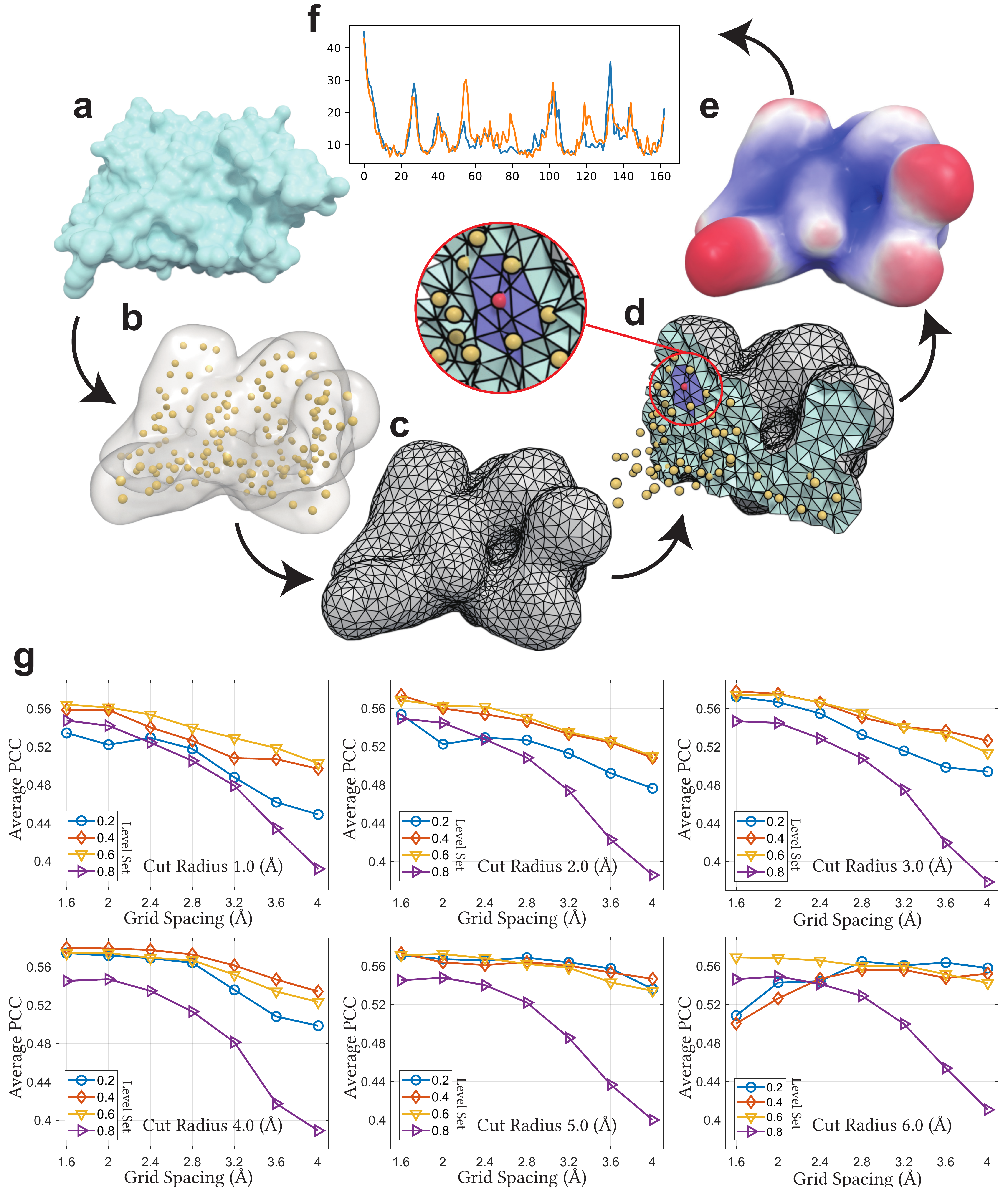}
	\caption{\textbf{Illustration of the procedure for flexibility analysis}. We use protein   3VZ9 as an example to demonstrate our procedure from \textbf{a} to \textbf{f}. \textbf{a} shows the input protein crystal structure. \textbf{b} shows that only C-alpha atoms (yellow spheres) are considered in this case. In fact, our method can analyze the flexibility of all atoms. We assign a Gaussian kernel to each C-alpha atoms and extract the level set surface (transparent surface) as our computation domain. \textbf{c} shows that standard tetrahedral mesh is generated with the domain (boundary faces are gray, inner faces are indigo.). We use a standard matrix diagonalization  procedure to obtain eigenvalues and eigen functions. B factor at each mesh vertice is computed as shown in Eq. (\ref{eqn:Bfactor}).
        \textbf{d}  B factor at the position of a C-alpha atom is obtained by the linear regression using within the nearby region    (For the red C-alpha atom, the linear regression region is colored as purple, which is within the cutoff radius.)
    \textbf{e} shows the predicted B factors on the surface. \textbf{f} shows the predicted B factors at C-alpha atoms (orange), compared with the experimental  B factors in the PDB  file (blue). Our prediction for 3vz9 has the Pearson correlation coefficient of 0.8081. \textbf{g} shows statistics of the average Pearson correlation coefficient (PPC) with various parameters on the test set of 364 proteins. Each plot has the same cutoff radius varying from 1.0 \AA~ to 6.0 \AA~ with interval 1.0 \AA. In each plot, the  level set value varies from 0.2 ~ to 0.8 ~ with interval 0.2 ~ shown by different lines; the grid spacing varies from 1.6 \AA~ to 4.0 \AA~ with interval 0.4 \AA~ shown in horizontal axis.}
	\label{fig:fluctuation}
\end{figure*}

We perform numerical experiments to confirm that our flexibility analysis is robust and reliable.   The cutoff radius is set to 7 \AA. Our method involves several parameters including, level set value $c$ and grid spacing $r$ and cutoff radius $d$ (See Fig. \ref{fig:fluctuation}).

\textbf{Level set} The level set  parameter   $c$ in Eq. (\ref{manifold}) controls the general distance from the surface to C-alpha atoms (See Fig. \ref{fig:fluctuation} (a)). A larger   level set value will result in a smaller  domain with richer topology structures, including many tunnels and cavities. While a smaller level set value will make the surface fatter so that it will lead to a ball-like shape.

\textbf{Grid spacing} The grid spacing $r$ controls the density of tetrahedrons of the mesh. A finer mesh will lead to a better prediction but is computationally more expensive (See Fig. \ref{fig:fluctuation} (b)).

\textbf{Cutoff radius} The parameter cutoff radius $d$ controls the linear regression region around the specific C-alpha atom \ref{fig:fluctuation} (d). Our approach will potentially introduce a denser mesh, which will lead to small local vibrations that should be filtered out. This treatment is the same as throwing away higher frequencies.

\begin{figure}
	\centering
	\includegraphics[width=.98\linewidth]{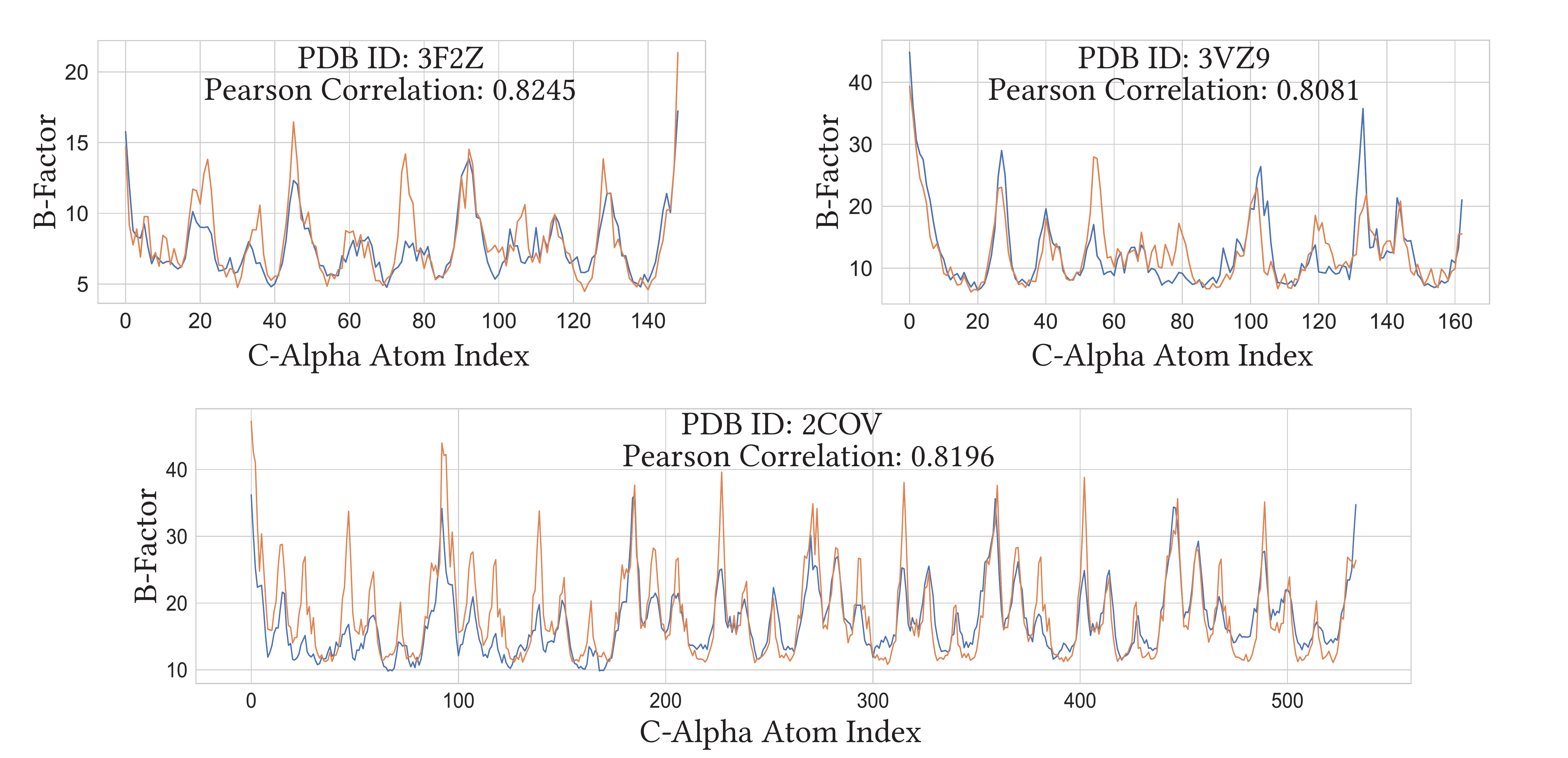}
	\caption{\textbf{Illustration of B-factor prediction.} We use proteins 3F2Z, 3VZ9 and 2COV as examples to show our predictions compared with the experiments. The blue lines are the ground truth from experiments. The orange lines are predictions from with our method.}
	\label{fig:bfactor}
\end{figure}
We consider a benchmark test set of 364 proteins studied in earlier work  \cite{Opron:2014} to systematically validate our method. Our test indicates that the best parameters are $c=0.4$, $r=1.6$ \AA, $d=4.0$ \AA. Fig. \ref{fig:bfactor} shows several examples with the best parameters. Table. \ref{tab:pctab} shows the average Pearson correlation coefficient of predicting the benchmark set of 364 proteins   \cite{Opron:2014} at a cutoff radius 4.0 \AA, which includes the overall best average Pearson correlation coefficient at grid spacing 1.6 \AA~ and level set value 0.4 .  The contour level value should not be too large such that only those C-alpha atoms that are close enough to each other will have interactions, as well as not be too small such that enough geometric and topological features are preserved. The cutoff radius should be a proper value such that higher frequencies are mitigated while lower frequencies are well kept. There is not much of influence of resolution if the previous 2 parameters are well set (see statistics at cutoff radius 5 \AA.). This provides the foundation for analyzing large protein complex with coarse resolution.

\begin{table}[]
\centering
\begin{tabular}{@{}|c|c||*{7}{c|}}  \hline
	\multicolumn{1}{|c}{}  &   &\multicolumn{7}{c|}{Grid Spacing (\AA~)}\\ \hline
	\multicolumn{1}{|c}{} & &1.6    &2.0    &2.4    &2.8    &3.2    &3.6    &4.0    \\ \hline \hline
	\multirow{4}*{\rotatebox{90}{Level Set}}
	& 0.2 &        0.574    &0.572    &0.569    &0.564    &0.536     &0.508    &0.498       \\ \cline{2-9}
	& 0.4 &\textbf{0.580}   &0.579    &0.578    &0.573    &0.561     &0.547    &0.534       \\ \cline{2-9}
	& 0.6 &        0.574    &0.574    &0.569    &0.567    &0.552     &0.534    &0.523       \\ \cline{2-9}
	& 0.8 &        0.545    &0.547    &0.535    &0.513    &0.481     &0.417    &0.389       \\ \hline
\end{tabular}
\caption{The average Pearson correlation coefficient for predicting 364 proteins at cutoff radius 4.0 \AA. The overall best average Pearson correlation coefficient is  0.580, compared to that of 0.565 for GNM on the same dataset \cite{Opron:2014}.
}
\label{tab:pctab}
\end{table}

The proposed  flexibility analysis can be easily extended to analyze the flexibility of a cryo-EM data at given level set.  The computed (relative) B factors are located at vertices but  can be interpolated to any desirable location if necessary. Due to the multi-resolution nature of our approach, computational cost is determined by the number of unknowns, i.e.,  the mesh size. For a given computational domain, the mesh size depends on the grid spacing. Therefore, for large macromolecules with millions of atoms, which is intractable for coordinate-based methods, the proposed de Rham-Hodge    approach can still be very efficient.

\paragraph{Natural mode analysis}

Normal mode analysis is an important approach for understanding biomolecular collective behavior, residue coupling, protein domain motion,  and protein-protein interaction,  reaction pathway, allosteric signaling, and enzyme catalysis  \cite{Go:1983,Tasumi:1982,Brooks:1983,Levitt:1985,JMa:2005}. However, normal model analysis becomes very expensive for large biomolecules. In particular, it is difficult to carry out the anisotropic network model (ANM) analysis \cite{Atilgan:2001} for cryo-EM maps which do not have atomic coordinates. Virtual particle-based ANM methods were proposed to tackle this problem \cite{tama2002exploring,ming2002describe}. Being based on the harmonic potential assumption, these methods are restricted to relatively small elastic motions.   In this work, we propose an entirely different strategy for biological macromolecular anisotropic motion analysis based on de Rham-Hodge theory.

\textbf{Laplace-de Rham operator}
It is noted that the mass-spring system is underlying many earlier successful elastic network models. This system describes the interconversion between the kinetic energy and potential energy during the dynamic motion. In our construction,  we take advantage of de Rham-Hodge theory. In fact, de Rham-Hodge theory provides a general framework to model the dynamic behavior of macromolecules. In the present work, we just illustrate this approach with special construction.

In order for  de Rham-Hodge theory to be able to describe anisotropic motions, we utilize the 1-form Laplace-de Rham operator
\begin{equation}\label{eqn:1-formDH}
\Delta_1 = d_0 \star_0^{-1} d_0^T \star_1 + \star_1^{-1} d_1^T \star_2 d_1,
\end{equation}
where $d_k$ denote   exterior derivatives on $\Omega^k (M)$ and $\star_k$ denote Hodge star operators.  Note that 2-form Laplace-de Rham operator works similarly well but we will limit our discussion with 1-form. The first term on the right hand side of Eq. (\ref{eqn:1-formDH}) is the quadratic energy form measuring the total divergence energy, while the second term  measures the total curl energy. Both terms are  kinetic energy physically or  Dirichlet energy mathematically.

\textbf{Laplace-de Rham-Helfrich operator}
Physically, a potential energy term is required to constrain the elastic motion of biological macromolecules.
There are many options, such as Willmore energy, which minimize the difference between two principle curvatures. Additionally, Helfrich introduced a curvature energy for modeling cell membrane or closed lipid vesicles \cite{Helfrich:1973,du2004phase}.  In our case, we assume the curvature energy of the form
\begin{equation}\label{eqn:1-Helfrich}
 V=\mu \int_{\partial M}(H-H_0)^2 dA,
\end{equation}
where $\mu$ is the molecular bending rigidity,  $H$ is the mean curvature on the molecular surface and $H_0$ is the spontaneous curvature of the molecule. The potential energy in Eq. (\ref{eqn:1-Helfrich}) is defined on the compact manifold enclosing a smooth molecular surface.
 
Conceptually, our curvature model deals with a dynamical system with a thin shell having thickness much smaller than other dimensions. Computationally, the 2D curvature model  serves as a boundary condition to complete  the Laplace-de Rham operator on a macromolecule.
The curvature energy increases as the mean curvature $H$ deforms away from its rest state. Therefore,  $H$ is a function of  the surface displacement.   The quadratic energy generated  from surface deformation is given by
\begin{equation}
Q = \partial^2 V / \partial X^2
\end{equation}
where  $X$ is  a displacement vector on the surface. Its  1-form representation can be expressed as
\begin{equation}
X = G \omega.
\end{equation}
Then the quadratic  form for the curvature energy in terms of  the 1-form is $G^T Q G$. Finally, the total 1-form quadratic energy is given by the following one-parameter Laplace-de Rham-Helfrich operator
\begin{equation}
E_{\mu}=   d_0 \star_0^{-1} d_0^T \star_1 + \star_1^{-1} d_1^T \star_2 d_1 +   G^T Q G
\end{equation}
We can solve the eigenvalue problem for the Laplace-de Rham-Helfrich operator  $E_{\nu}$ to extract the natural vibration modes of biomolecules. It is a standard procedure to assemble required matrix $G$ and $Q$
together with our Laplace-de Rham matrix.

\begin{figure}
	\centering
	\includegraphics[width=.98\linewidth]{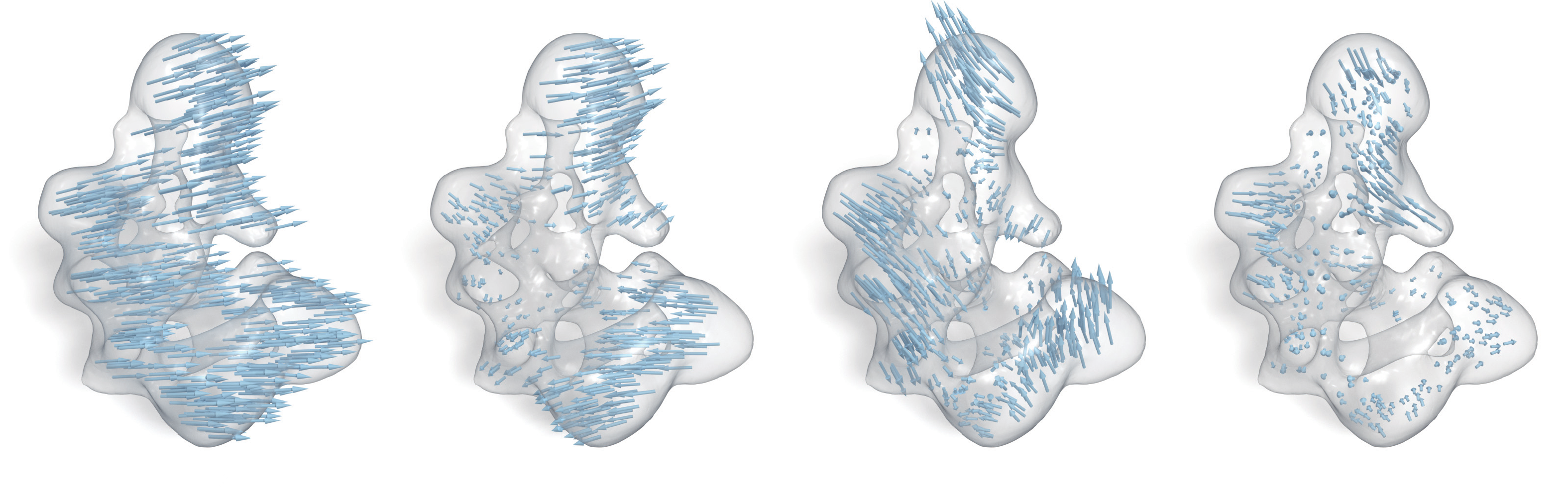}
	\caption{\textbf{Natural modes of EMD 1258.} The 0-th, 4-th, 8-th and 12-th natural modes are shown. }
	\label{fig:modal}
\end{figure}

In fact, an advantage of the proposed anisotropic motion theory is that it allows to treat the divergence energy and curl energy differently. For example, we can introduce a bulk modulus-type of parameter
$\lambda$ to the divergence energy term, which leads to a  weighted Laplace-de Rham  operator. As a result, we have a two-parameter Laplace-de Rham-Helfrich operator
\begin{equation}
E_{\lambda\nu}= \lambda\cdot  d_0 \star_0^{-1} d_0^T \star_1 + \star_1^{-1} d_1^T \star_2 d_1 +   G^T Q G.
\end{equation}
 We need to choose appropriate weight parameters $\lambda$ and $\mu$. Generally, the two-parameter Laplace-de Rham-Helfrich operator and boundary condition matrix can be tuned separately.  
What we would like to achieve is letting the curvature energy drive the motion and let our system penalize the compressibility (i.e., the divergence energy).  Therefore,  we select an appropriate  $\lambda$ at a different scale and choose  $\mu > \lambda > 1$.
 
Modal analysis, compared to fluctuation analysis, provides   more information. In addition to  the description of flexibility, modal analysis also provides the collective motion of a molecule and  its potential  function.  The dynamics of a macromolecule can be described by the linear combination of its natural  modes. Fig. \ref{fig:modal} shows several natural modes for core spliceosomal components, EMD 1258, which indicates the success of our    Laplace-de Rham-Helfrich operator.

It is noted that the original Laplace-de Rham operator with appropriate boundary conditions admits the orthogonal Hodge decomposition in terms of divergence-free, curl-free, and harmonic eigenmodes. In contrast,  the  Laplace-de Rham-Helfrich operator does preserve these properties. Nonetheless,   the eigenmodes generated by the Laplace-de Rham-Helfrich operator are mutually orthogonal and subject to different physical interpretations.   For example, the first three eigenmodes are associated with 3D translational motions. The modes in Fig. \ref{fig:modal} have little to do with the topological singularity of  EMD 1258.

Additionally, the eigenmodes in Fig. \ref{fig:L1t} have a fixed boundary. In contrast, boundaries of eigenmodes generated with the Laplace-de Rham-Helfrich operator as shown in Fig. \ref{fig:modal} are allowed to change. The Laplace-de Rham-Helfrich operator can predict significant macromolecular deformations, which are controllable with two weight parameters, $\lambda$ and $\mu$. In contrast,  existing normal mode analysis methods can only admit small deformations due to the use of the harmonic potential.

Finally, due to its continuous nature, the proposed Laplace-de Rham-Helfrich operator can be easily employed for natural mode analysis at any given scale. It can be directly applied to the analysis of cryo-EM maps and other volumetric data at an arbitrary scale. One specific example of potential applications is the analysis of subcellular organelles, such as mitochondrial ultrastructure and endoplasmic reticulum.

\section{Methods}\label{Sec:Methods}

We provide the details for our design of computational tools, data structures, and parameters in our implementation of the present de Rham-Hodge spectral analysis. Through efficient implementation, our method is highly scalable and capable of handling molecular data ranging from protein crystal structures to cryo-EM maps.

\subsection{Simplicial complex generation}

The domain of our Laplace-de Rham operators is first tessellated into a simplicial complex, which is a tetrahedral mesh in our 3D case. There are quite a few well-developed software packages for tetrahedral mesh generation given a boundary with a surface triangle mesh as input. We chose CGAL (Computational geometry algorithm library) over others for its superior control on element quality.

In theory, we can generate tetrahedral meshes with any highly accurate closed surface. However, macromolecule complexes with atom-level resolution often make the output mesh intractable with typical computing platforms. Moreover, a dense mesh is unnecessary for the calculation of the low-frequency range of the spectrum. Thus, we produce a coarse resolution with a spatial sampling density higher than twice the spatial frequencies (wavenumbers) of the geometrical and topological features to be computed in the given biomolecule complexes.

For protein crystal structures, we tested the construction of the surface using only the $C_\alpha$ positions. First, a Gaussian kernel is assigned to each atomic position to approximate the electron density. Then, a level set surface is generated to construct the contour of the protein closely enclosing the high electron density regions.

For cryo-EM data, to produce a smooth contour surface, Gaussian kernels are associated with data points. Other approaches, such as mean curvature flow \cite{Bates:2008, RDZhao:2018a} can be used as well. When dealing with noisy and densely sampled data, we can carefully choose the level set that corresponds to a fairly smooth contour surface that encloses the original cryo-EM data.

Given a volumetric data, we can either directly use CGAL to produce a tetrahedral mesh, or first convert it to a triangular surface mesh through the marching cubes algorithm, and use that to generate a tetrahedral mesh. Different sampling densities are tested to meet typical quality requirements while balancing computational cost and mesh quality.

\subsection{Discrete exterior calculus}

As a topological structure-preserving discretization of the exterior calculus on differential forms, discrete exterior calculus (DEC) has been widely applied in the recent years for various successful applications on geometrical problems and finite element analysis, including meshing and computational electromagnetics. It is an appropriate tool for our de Rham-Hodge analysis of biomolecules, as all the related operations, including exterior derivatives and the Hodge stars, are represented as matrices that preserve the defining properties in the continuous setting. More precisely, the discrete exterior derivative operators strictly satisfies $D_{k+1} D_{k}=0$, mimicking $d_{k+1} d_k=0$, and the discrete Hodge star operators are realized by symmetric positive definite matrices. Hence, the discrete Laplace-de Rham operators can be assembled using finite dimensional linear algebra with the aforementioned three distinct spectra.

\begin{figure}
	\centering
		\includegraphics[width=.98\linewidth]{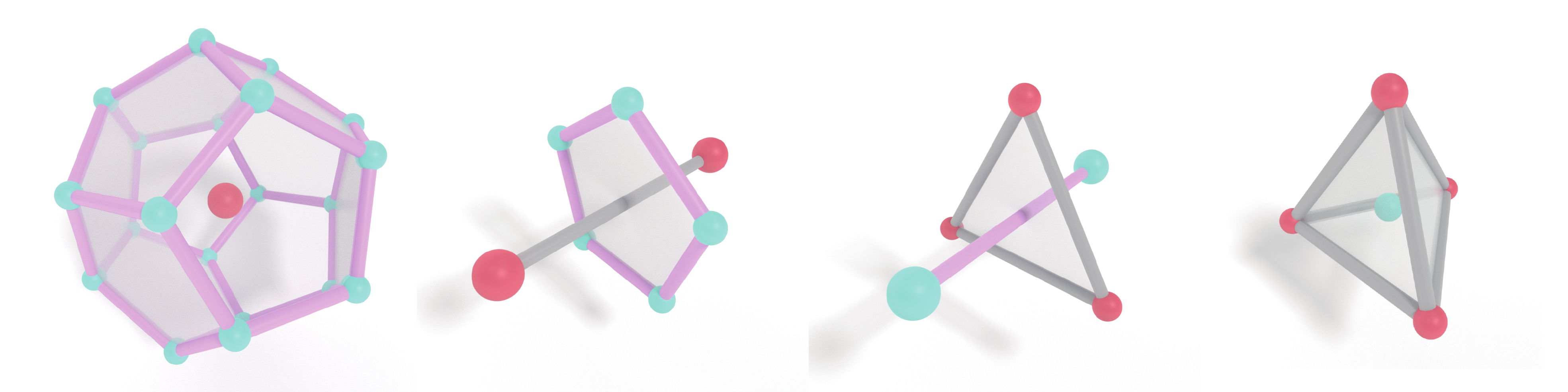}
	\caption{\textbf{Illustration of the primal and dual elements of the tetrahedral mesh}.
	All the red vertices are mesh primal vertices.  All the indigo vertices are dual vertices at circumcenter of each tet. All the gray edges are primal edges. All the pink edges are dual edges connecting adjacent dual vertices.
The first chart shows the dual cell of a primal vertex.
The second  chart shows the dual facet of the primal edge.
The third chart shows the dual edge of the primal facet.
The last chart shows the dual vertex of the primal cell (tet).
}
	\label{fig:primaldual}
\end{figure}

To allow replication of our results, we recap our implementation of DEC in the following. We start by a tetrahedral tessellation of the volumetric domain, i.e., a tetrahedral mesh, which is the collection of a vertex set $\mathcal{V}$, an edge set $\mathcal{E}$, a triangle set $\mathcal{F}$, and a tetrahedron set $\mathcal{T}$. The vertices are points in 3D Euclidean space, the edges/triangles/tetrahedra are represented as $1$-/$2$-/$3$-simplices, i.e., pairs/triples/quadruples of vertex indices respectively, and regarded as the convex hull of these vertices. We further choose an arbitrary orientation for each $k$-simplex, which is an order set of $k\!+\!1$ vertices, up to an even permutation. We denote an oriented $k$-simplex as
\begin{equation}
	[\sigma] = [v_1, v_2, ..., v_k].
\end{equation}
The boundary operator is defined as
\begin{equation}
	\partial [\sigma] = \sum_{i=1}^{k} [v_1, v_2, ..., \hat{v}_i,..., v_k],
\end{equation}
where $\hat{v}_i$ means that the $i$-the vertex is omitted. Thus the boundary operator will take all the $1$-degree lower faces of $\sigma$ with an induced orientation. We will take the following strategy to handle orientation in the implementation. We usually assign each tet an orientation such that, when applying the boundary operator, each facet has an outward pointing orientation. The total boundary of the tet mesh conforms naturally with the surface with outward pointing orientation. But for each edge and facet, we pre-assign an orientation by increasing indices of incident vertices. In this case, we need to take care of the boundary operator when there is a conflict between the pre-assigned orientation and the induced orientation.

\begin{figure}
	\centering
	\includegraphics[width=.8\linewidth]{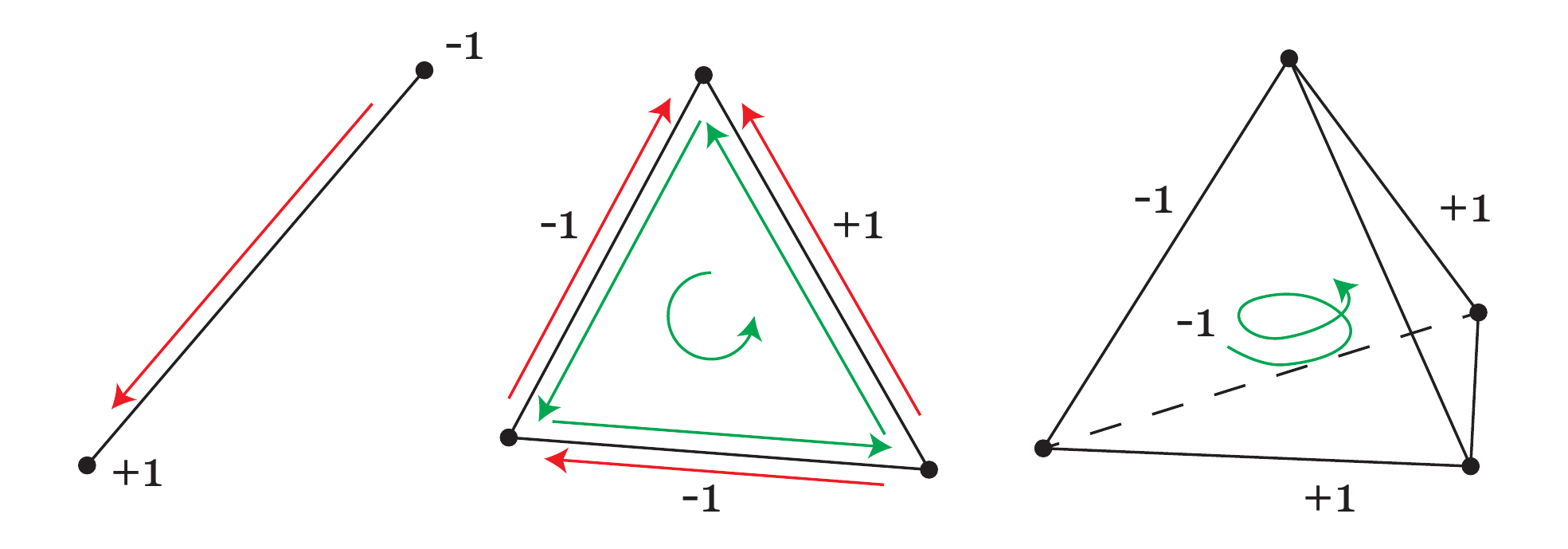}
	\caption{\textbf{Illustration of orientation}. Pre-assigned orientation is colored in red. Induced orientation by $\partial$ is colored in green. The vertices are assumed to have a positive pre-assigned orientation. Therefore, the induced orientation from edge orientation is $+1$ at the head and $-1$ at the tail. For a triangle facet, $+1$ is assigned whenever the pre-assigned orientation conforms with the induced orientation, and $-1$ vice versa. A similar rule will apply to tets. }
	\label{fig:orientation}
\end{figure}

Scalar fields are naturally encoded as 0-form and 3-form. A 0-form is the same with the finite element method with each vertex sampled a value. A 3-form is, different from a 0-form, stored per tet as volume integration of the scalar field. Vector fields are naturally encoded as 1-form and 2-form. A 1-form is sampled by the line integral on each oriented edge. A 2-form is sampled by surface flux on each oriented facet. Whitney forms \cite{bossavit1988whitney} can help convert forms back to piece-wise constant vector fields on each tet.

\begin{figure}
	\centering
	\includegraphics[width=.70\linewidth]{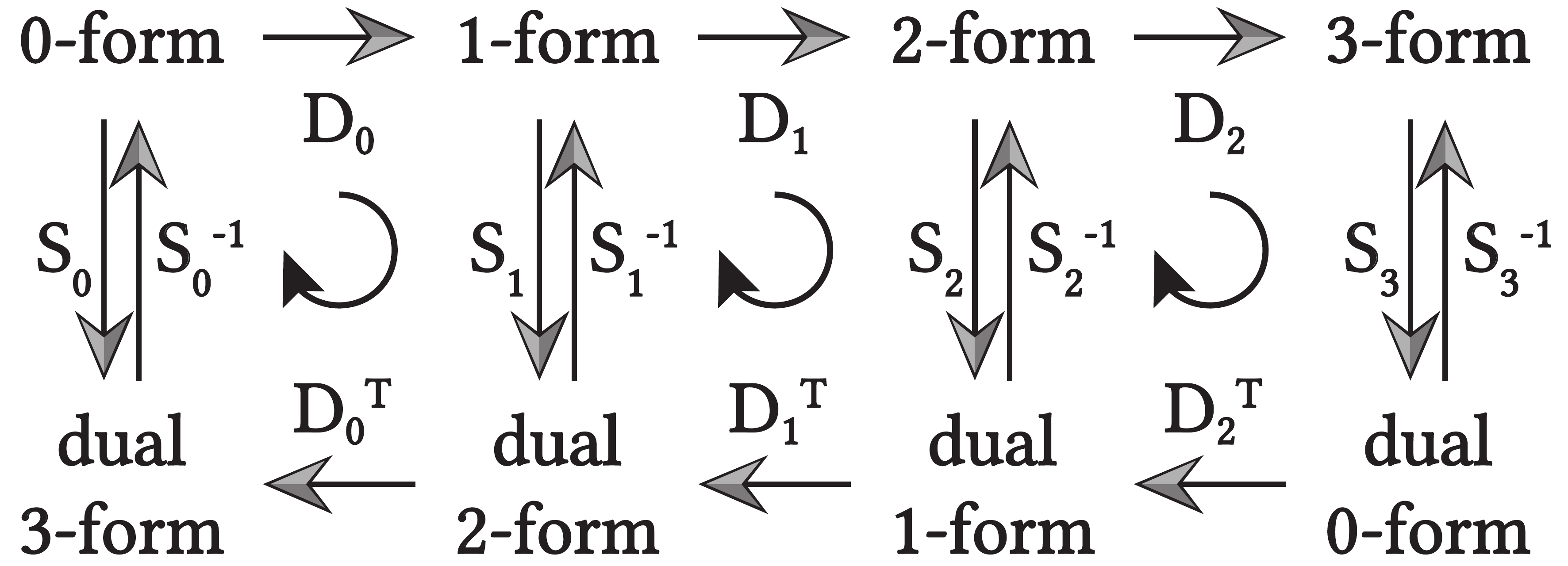}
	\caption{\textbf{Illustration of cohomology}. This figure illustrates the relation by exterior derivative and Hodge star operators. The assembly of Laplacian operator $L_k$ is just starting from primal $k$-forms, multiplying matrices along the circular direction. }
	\label{fig:cohomology}
\end{figure}

We will store discrete $k$-forms as column vectors. Then as mentioned before, all the discrete operators can be formed as matrices applying on the column vectors. Then we start to construct discrete exterior derivative and discrete Hodge star matrices. Suppose we are dealing with discrete differential form $d\omega$ on simplices $\sigma$, according to Stokes' theorem
\begin{equation}
\int_{\partial \sigma} \omega = \int_{\sigma} d \omega,
\end{equation}
$d\omega$ is just an oriented summation of $\omega$ on facets of $\sigma$. So the discrete exterior derivative operator $D_k$ is just a matrix filled with $-1, 0, 1$ (see Fig. \ref{fig:orientation}), depending on whether the pre-assigned orientation is conforming with the induced orientation. One can easily observe that the discrete exterior derivative operators for dual forms are merely $D_k^{T}$. The discrete Hodge star operator $S_k$ is just converting primal form and dual form back a forth.  Each primal element in the tet mesh has one corresponding dual element (See Fig. \ref{fig:primaldual}). So the discrete Hodge star operator is merely a diagonal matrix. Note that here we use a diagonal matrix to approximate the Hodge star operator, where non-diagonal Hodge star with higher accuracy can be applied as well. But a diagonal Hodge star is enough for our current application. The diagonal Hodge star matrix just has diagonal entries as dual element volume over primal element volume. For example, given a 1-form on each edge, applying the Hodge star is turning the primal 1-form into dual 2-form stored on each dual facets. This can be interpreted as we sample the vector field at the center of the edge. One way is to compute the 1-form as the sampled vector integrated the primal edge as the line integral, the other way is to compute the 2-form as the sampled vector integrated on the dual facet as vector flux. So the transition can be encoded as a number dual element volume over primal element volume. See Fig. \ref{fig:cohomology} for relations between differential forms and operators.

\begin{figure}
	\centering
	\includegraphics[width=.30\linewidth]{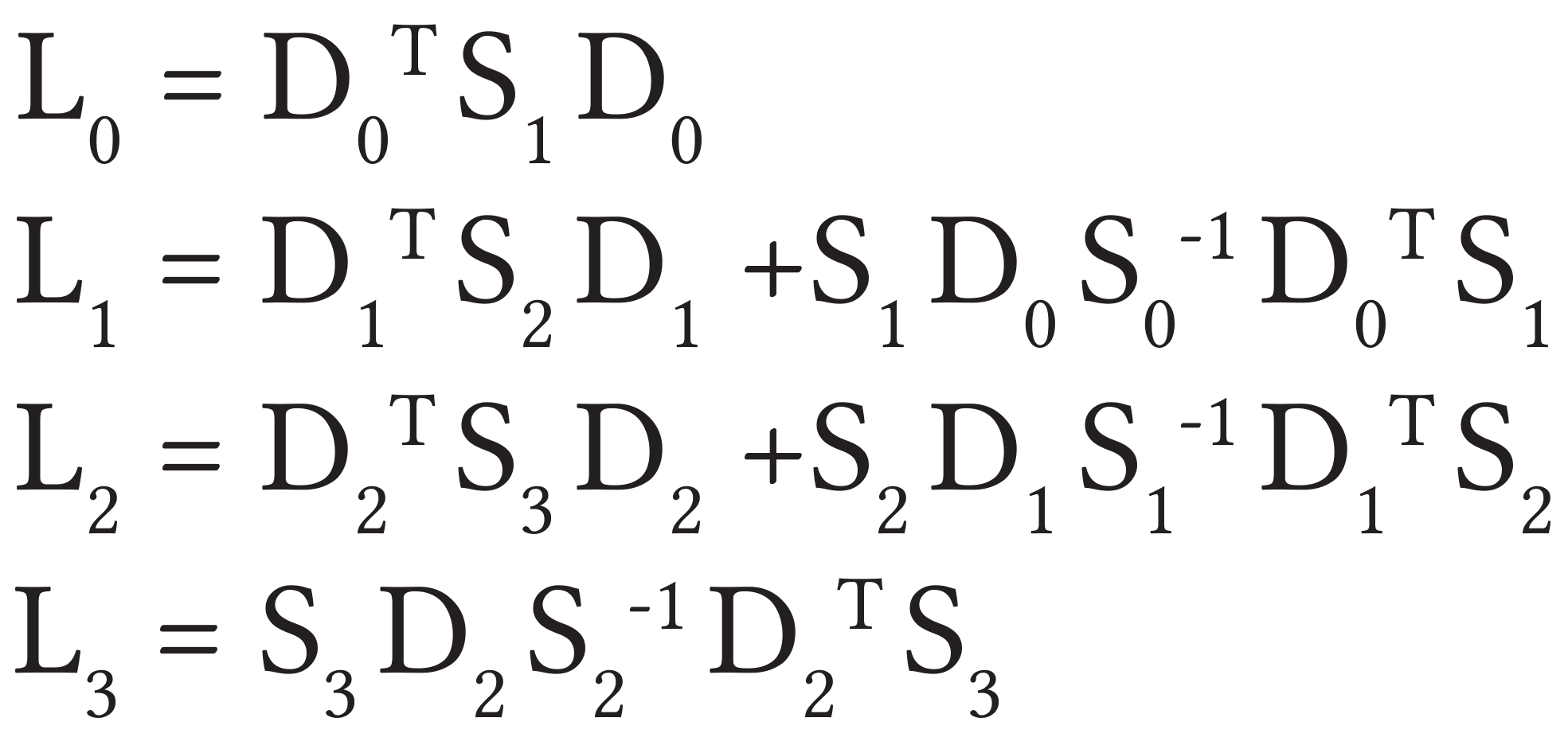}
	\caption{\textbf{Illustration of Laplacian assembly}. Laplacian matrix is assembled by matrix multiplication with pre-assembled discrete exterior derivative and discrete Hodge star matrices.}
	\label{fig:laplacian}
\end{figure}

Once we have these related matrices for discrete operators, we are ready to construct the Laplacian matrix $L_k$ according to Fig. \ref{fig:laplacian}. The assembly of Laplace-de Rham operator $L_k$ is just starting from primal $k$-forms, multiplying matrices along the circular direction as shown in Fig. \ref{fig:laplacian}. Note that the usual Hodge Laplacian matrix is not symmetric generally. In practice, we usually left multiply by Hodge star to turn it into a symmetric one. After this, we need to take care of the boundary conditions. Boundary condition treatment can be incorporated when assembling $d$ matrices. Recall that the $d$ matrices are merely for creating an oriented summation of discrete differential forms stored on simplices. We can just delete corresponding columns and rows for boundary elements. We use $L_{k,t}$ to denote Laplace-de Rham operator with boundary elements, and $L_{k,n}$ to denote those without boundary elements \cite{demlow2014posteriori}.

Finally, the spectral analysis can be done with a generalized eigenvalue problem.
\begin{equation}
L_k v = \lambda S_k v
\end{equation}
The leading small eigenvalues and corresponding vectors are associated with useful low frequencies. In principle, large eigenvalues also contain useful information but are often impaired by large computational errors.   We use an eigensolver with parameter starting from small magnitude eigenvalues.

\section{Conclusion}

The de Rham-Hodge theory is a landmark of 20th Century’s mathematics that interconnects differential geometry, algebraic topology, and partial differential equation. It provides a solid mathematical foundation to electromagnetic theory, quantum field theory and many other important physics. However,  this important mathematical tool has never been applied to macromolecular modeling and analysis, to the best of our knowledge. This work introduces the de Rham-Hodge theory as a unified paradigm to analyze biomolecular geometry, topology,  flexibility and natural mode based on three-dimensional (3D) coordinates or cryo-EM maps.  Specifically, de Rham-Hodge spectral analysis has been carried out to reveal macromolecular geometric characteristic and topological invariants with normal and tangential boundary conditions. The Helmholtz-Hodge decomposition is employed to analyze the divergence-free, curl-free, and harmonic components of macromolecular vector fields. Based on the 0-form scalar Hodge-Laplacian,  an accurate multiscale model is constructed to predict protein fluctuations.  By equipping a vector Laplace-de Rham operator with a boundary constraint based on Helfrich-type curvature energy, a 1-form Laplace-de Rham-Helfrich operator is proposed to predict the natural modes of biomolecules, particularly cryo-EM maps. In addition to its versatile nature for a wide variety of modeling and analysis, the proposed de Rham-Hodge paradigm also provides a unified approach to handle biomolecular problems at various spatial scales and with different data formats.  A state-of-the-art 3D discrete exterior calculus algorithm is developed to facilitate accurate, reliable and topological structure preserving spectral analysis and modeling of biomolecules. Extensive numerical experiments indicate that the proposed de Rham-Hodge paradigm offers one of the most powerful tools for the modeling and analysis of biological macromolecules.

The proposed de Rham-Hodge paradigm provides a solid foundation for a wide variety of other biological and biophysical applications. For example, the present de Rham-Hodge flexibility and natural mode analysis can be directly applied to subcellular organelles, such as vesicle,  endoplasmic reticulum, Golgi apparatus, cytoskeleton,   mitochondrion, vacuole, cytosol,  lysosome, and centrosome, for which the existing atomistic biophysical approaches have very limited accessibility. Additionally, features extracted from de Rham-Hodge flexibility and natural mode analysis can be incorporated into deep neural networks for the structure reconstruction from medium and low-resolution cryoEM maps \cite{haslam2018exploratory}. Finally, due to its ability to characterize geometric traits and describe topological invariants, the proposed de Rham-Hodge paradigm opens an entirely new direction for the quantitative structure-function analysis of molecular and macromolecular datasets. The integration of de Rham-Hodge features and machine learning algorithms for the predictions of protein-ligand binding affinity, protein-protein binding affinity, protein folding stability change upon mutation, drug toxicity, solubility, partition coefficient, permeability, and plasma protein binding are under our consideration.

 \section{Acknowledgments}
This work was supported in part by  NSF Grants DMS-1721024, DMS-1761320, and IS1900473,  and NIH grant  GM126189.  GWW was also funded by Bristol-Myers Squibb and Pfizer.



\end{document}